\documentclass[12pt,a4paper,aps,preprint,superscriptaddress,nofootinbib]{revtex4-1}
\usepackage[utf8]{inputenc}
\usepackage{graphicx}
\usepackage{amssymb}
\usepackage{textcomp}
\usepackage{amsmath}
\usepackage{tabularx}
\usepackage{bm}
\usepackage{times}
\usepackage{color}
\usepackage{slashed}
\usepackage{multirow}
\usepackage{verbatim}
\usepackage{cancel}
\usepackage{subfigure}
\usepackage{ulem}
\usepackage{float}
\newcommand{\ndens}{\mathfrak{n}}  
\newcommand{\ntarget}{\mathcal{N}}  

\usepackage[colorlinks=true, pdfstartview=FitV, linkcolor=blue, citecolor=blue, urlcolor=blue]{hyperref}

\allowdisplaybreaks[4]

\def\lsim{\mathrel{\raise.3ex\hbox{$<$\kern-.75em\lower1ex\hbox{$\sim$}}}}
\def\gsim{\mathrel{\raise.3ex\hbox{$>$\kern-.75em\lower1ex\hbox{$\sim$}}}}

\definecolor{orange}{rgb}{1,0.5,0}

\begin{document}

\title{Constraints on light dark matter from primordial black hole evaporation at dark matter direct detection experiments}

\author{Tong Zhu}
\email{zhut55@mail2.sysu.edu.cn}
\affiliation{
School of Physics, Sun Yat-Sen University, Guangzhou 510275, China
}

\author{Cheng-Rui Jiang}
\email{jiangcr@mail.nankai.edu.cn}
\affiliation{
School of Physics, Nankai University, Tianjin 300071, China
}

\author{Tong Li}
\email{litong@nankai.edu.cn}
\affiliation{
School of Physics, Nankai University, Tianjin 300071, China
}

\author{Jiajun Liao}
\email{liaojiajun@mail.sysu.edu.cn}
\affiliation{
School of Physics, Sun Yat-Sen University, Guangzhou 510275, China
}

\begin{abstract}
Primordial black holes (PBHs) are able to produce light dark matter (DM) particles via Hawking radiation, and yield a flux of boosted DM that can be probed at underground DM direct detection experiments. We analyze both galactic and extragalactic contributions to the differential flux of light DM from PBH evaporation, and then compute the expected event rate from PBH boosted DM scattering off electrons or nuclei after taking into account the attenuation effect. Using recent data from DM direct detection experiments XENONnT, PandaX-4T and LZ, we set constraints on both DM--electron and DM--nucleus scattering cross sections, as well as the fraction of DM composed of PBHs $f_{\rm PBH}$ for $9\times10^{14}\text{--}1\times10^{16}\,\mathrm{g}$ PBHs that are not fully evaporated today. We also investigate the spectral evolution induced by Hawking evaporation throughout the evaporation and post-evaporation regimes. The constraints on the PBH mass are then extended into the $1\times10^{13}\text{--}6\times10^{14}\,\mathrm{g}$ window for fully evaporated PBHs.
\end{abstract}

\maketitle

\tableofcontents

\newpage

\section{Introduction}
\label{sec:Intro}

The existence of dark matter (DM) is strongly supported by astrophysical and cosmological observations~\cite{Zwicky:1933gu,Rubin:1970zza,Clowe:2006eq,Planck:2018vyg}. However, its underlying physical nature remains elusive.
A well-known possibility is the weakly interacting massive particle (WIMP), which naturally arises in many extensions of the Standard Model (SM), and can reproduce the observed relic abundance through thermal freeze-out scenario~\cite{Lee:1977ua,Griest:1990kh,Bertone:2010at}. However, DM direct detection (DD) experiments have not yet observed a signal of WIMP, pushing the upper limit of DM--nucleon scattering cross section down to $\sim 10^{-46}$–$10^{-47}\,\text{cm}^2$~\cite{XENON:2025vwd,PandaX:2024qfu,LZ:2024zvo}. These null results have motivated the exploration of alternative DM scenarios, such as sub-GeV light DM. Conventional WIMP detectors rapidly lose sensitivity for DM below GeV scale due to suppressed nuclear recoil energies. However, light DM particles with masses in the keV--GeV range originating from a hidden sector may evade WIMP constraints while remaining detectable through new strategies~\cite{Pospelov:2007mp,Boehm:2003hm,Pospelov:2008jk}. Recent progress, including the most recent results from XENONnT~\cite{XENON:2024znc}, SENSEI~\cite{SENSEI:2023zdf}, PandaX~\cite{PandaX:2023xgl}, CDEX~\cite{CDEX:2022kcd} and DarkSide~\cite{DarkSide:2022dhx}, has opened new parameter space for sub-GeV DM, complementary to traditional WIMP searches.

Another compelling possibility of DM beyond WIMP is primordial black holes (PBHs)~\cite{Zeldovich:1967lct,Carr:1974nx,Carr:1975qj,Khlopov:2008qy,Belotsky:2014kca,Carr:2021bzv}. They can be produced by the collapse of density perturbations in the early Universe. Unlike astrophysical black holes, PBHs can span a wide mass range, potentially as small as $10^{12}$–$10^{16}$\,g. Although observational constraints have ruled out PBHs as the dominant DM component across many mass ranges, the observation of $\sim 30\,M_\odot$ black hole mergers at LIGO has renewed interest in the PBH DM~\cite{LIGO:2017dbh,Carr:2020gox}. Later studies indicate that PBHs of asteroid mass and certain supercluster mass could still constitute all of the DM without violating current bounds~\cite{Montero-Camacho:2019jte,Carr:2020erq,Carr:2020gox}. Therefore, both light particle DM and PBHs remain viable and are actively explored in the post-WIMP era~\cite{Carr:2009jm,Carr:2020gox,Auffinger:2022khh}.

An intriguing scenario connecting these two ideas is the production of light DM particles from PBHs via Hawking evaporation~\cite{Hawking:1975vcx,Page:1976df,Page:1977um,MacGibbon:1990zk,MacGibbon:1991tj,Khlopov:2020vpx,Khlopov:2024nqp,Friedlander:2023qmc}. Hawking radiation, arising from quantum effects near the event horizon, causes black holes to emit particles with a characteristic Hawking temperature inversely proportional to their mass. Thus, a non-rotating PBH with $M_{\rm PBH} \sim 10^{15}$ g has a temperature of $T_{\rm PBH} \sim \mathcal{O}(10)$\,MeV and allows the emission of all SM particles lighter than this scale. If DM consists of sub-GeV particles with $m_\chi \lesssim T_{\rm PBH}$, they may also be emitted via this process. These emitted DM particles carries kinetic energies $\sim T_{\rm PBH}$, significantly higher than the mass of cold halo DM.
Such light particles can propagate through the galaxy and generate observable signatures in underground detectors.

This Hawking evaporation mechanism provides a concrete example of boosted DM (BDM), where DM particles gains significant kinetic energy through decay or evaporation of heavier sources. Recent studies have explored the sensitivity of such PBH boosted DM (PBHBDM) at DM DD experiments, originally designed for non-relativistic WIMPs ~\cite{Calabrese:2021src,Li:2022jxo,Calabrese:2022rfa,Marfatia:2022jiz}. In particular, elastic scattering of evaporated DM particles with electrons has been searched at CDEX~\cite{CDEX:2024xqm}.
These works have shown that DM DD detectors and neutrino detectors can set competitive constraints on the DM--electron or DM--nucleus scattering cross section for sub–GeV DM, assuming PBHs consist of a subdominant fraction of the total DM density.

Despite these advances, significant gaps remain. In particular, recent analyses have not yet incorporated results from latest DM DD data, which includes larger exposures and improved background rejection, thereby can substantially enhancing the sensitivity to DM. In addition, most existing studies in the literature focus on one single detection mode, either DM--electron or DM--nucleus scattering. Realistic models may allow both of them, depending on the nature of DM couplings. Thus, a comprehensive consideration of both detection modes is needed. Furthermore, the modeling of the PBH population requires further attention. Most previous works consider a monochromatic mass distribution of PBHs and assume that PBHs have not fully evaporated. In reality, PBHs likely span a broad mass range, and those with masses below the evaporation mass threshold $M_{\rm th}\approx 7.5 \times 10^{14}~{\rm g}$ would have completely evaporated by now. The injection of stable particles, including DM, from these evaporated PBHs can be constrained by cosmological observations such as big bang nucleosynthesis and the extragalactic gamma ray background~\cite{Carr:2020gox,Arbey:2019vqx,Chen:2021ngo,Wu:2025ovd}. At the same time, these fully evaporated PBHs may contribute to a residual DM flux observable today. However, current studies have not jointly considered DM contributions from both actively evaporating and fully evaporated PBHs.

In this work, we address these gaps by performing a comprehensive analysis of light DM production via PBH Hawking evaporation using the latest data from DM detectors XENONnT~\cite{XENON:2022ltv}, PandaX-4T~\cite{PandaX:2024cic} and LUX-ZEPLIN (LZ)~\cite{LZ:2022lsv}. XENONnT is a direct DM search experiment located approximately $1.4\,\mathrm{km}$ underground at the Gran Sasso laboratory in Italy. The apparatus mainly consists of a dual-phase liquid–gas xenon time-projection chamber (TPC) surrounded by neutron and muon veto systems. Its active target comprises $5.9\,\mathrm{t}$ of liquid xenon (LXe), optimized for detecting low energy interaction signatures~\cite{XENON:2024wpa}. PandaX-4T, sitting at the China Jinping Underground Laboratory at a depth of about $2.4\,\mathrm{km}$, adopts a similar TPC-centered design with a total of $3.7\,\mathrm{t}$ LXe target~\cite{PandaX:2018wtu,Qian:2025kxb}. LZ, located at the Sanford Underground Research Facility in South Dakota, USA at a depth of roughly $1.478\,\mathrm{km}$, is another large-scale LXe TPC experiment searching for rare DM interactions~\cite{LZ:2019sgr}. Using the latest experimental data, we derive updated constraints on the DM--electron and spin-independent (SI) DM--nucleon scattering cross sections, as well as the PBH abundance parameter. We consider a general interaction framework in which DM couples to electrons or nucleons, and our analysis includes both interactions. Here we study either nucleus scattering or electron scattering which is independent of any realistic dark matter model. In particular, we follow Ref.~\cite{Barker:2012ek} to convert the nuclear recoil spectrum into electronic equivalent signals by using the quenching factor. Therefore we can impose constraints on the cross sections of both the nucleus scattering and electron scattering processes from the same electronic data at the DM DD experiments.

We compute the expected DM flux and recoil spectra for a wide range of PBH masses.
Importantly, we consider two scenarios for the PBH mass ranges. The first scenario corresponds to partially evaporating PBHs with $M_{\rm PBH}\in[9\times10^{14},\,10^{16}]\,\mathrm{g}$, which yields an ongoing local flux at Earth. The second one corresponds to fully evaporated PBHs with $M_{\rm PBH}<6\times10^{14}\,\mathrm{g}$, whose past emission may contribute to a relic DM background today. We only consider the monochromatic mass distribution at the production of PBHs in our work. Other types of mass distribution have been studied in Refs.~\cite{Carr:2017jsz,Bernal:2022swt}. For the second scenario, we further investigate the impact of PBH evolution for those with lifetimes shorter than the age of the Universe. By comparing the predicted event spectra in both scenarios with the null results from three state-of-the-art DD experiments, we derive stringent exclusion limits on the scattering cross section as a function of light DM mass.
We further convert these bounds into updated constraints on the fraction of DM composed of PBHs today $f_{\rm PBH}$ and the initial PBH abundance parameter $\beta'_{\rm PBH}$ for partially and fully evaporated PBHs, respectively. Our results significantly improve the previously constraints on the PBHBDM parameter space and highlight the complementarity of cosmological bounds and DD searches in probing such non-standard DM scenarios.

This paper is organized as follows. In Sec.~\ref{sec:PBHBDM}, we introduce the PBH evaporation mechanism and compute the resulting DM flux from PBH evaporation. The attenuation effect from Earth matter is also discussed. We calculate the predicted events of PBHBDM in Sec.~\ref{sec:scattering} and obtain the constraints from the latest data in DM DD experiments in Sec.~\ref{sec:results}. In Sec.~\ref{sec:evolution} we discuss the scenario of mass evolution of PBHs during evaporation. We summarize our main conclusions in Sec.~\ref{sec:conclusions}.

\section{Light DM from PBH evaporation}
\label{sec:PBHBDM}

In this section, we first introduce the PBH evaporation mechanism. The fermionic DM flux from PBH evaporation is then computed by taking into account the attenuation effect from Earth matter.

\subsection{PBH evaporation and light DM spectrum}
\label{sec:PBH}

Primordial black holes can be produced through a variety of mechanisms. Different formation scenarios generally yield different PBH mass distributions~\cite{Carr:2019kxo,Carr:2017jsz,Green:2016xgy,Liu:2021svg}. They can in turn modify the expected flux of light DM~\cite{Bellomo:2017zsr,Suyama:2019npc,Mosbech:2022lfg}. In this work, we adopt a monochromatic mass distribution for simplicity. We also neglect accretion for the mass evolution in this work since accretion is inefficient and does not lead to a significant mass growth~\cite{Carr:1974nx,Custodio:1998pv,Mack:2006gz}. We will first focus on PBH masses greater than about $10^{15}~{\rm g}$ and neglect PBH mass evolution , which will be discussed in Sec.~\ref{sec:evolution}.

We assume that there exists a new Dirac type of sub-GeV fermionic DM particle with spin $1/2$ beyond the SM. Although the inclusion of this new particle could in principle affect the evaporation rate of PBHs, under the assumption of extremely light DM, this effect is negligible. The differential spectrum of DM  particle, denoted by $\chi$, emitted via Hawking radiation is given by~\cite{Hawking:1975vcx}
\begin{equation}
\frac{d^2N_\chi}{dT_\chi\,dt} = \frac{g_\chi}{2\pi}\frac{\Gamma\left(T_\chi,M_{\mathrm{PBH}}\right)}{\exp\!\Bigl(\frac{T_\chi+m_\chi}{k_{\mathrm{B}}T_{\mathrm{PBH}}}\Bigr)+1}\,,
\label{eq:hawking_radiation}
\end{equation}
where $T_\chi$ is the kinetic energy of the DM particle $\chi$, $m_\chi$ is the DM mass, $t$ is the emission time, $g_\chi=4$ denotes the degree of freedom for Dirac DM particle, $M_{\rm PBH}$ is the PBH mass, $\Gamma\left(T_\chi,M_{\mathrm{PBH}}\right)$ is the graybody factor which modifies the blackbody spectrum, $k_B$ is Boltzmann's constant, and $T_{\mathrm{PBH}}$ is the Hawking temperature of the PBH. For the simulation of DM production, we employ the latest version of the \textbf{BlackHawk v2.3} code~\cite{arbey_blackhawk_2019,arbey_physics_2021}.

For the PBH--emitted particle spectra, the total flux is typically divided into two types of contribution. One contribution is from PBHs within the Milky Way (MW) and the other is from extragalactic (EG) PBHs~\cite{Bernal:2022swt,Li:2022jxo}. For PBHs with mass above the evaporation threshold $M_{\rm th}$ $\approx 7.5 \times 10^{14}~{\rm g}$, we use \textbf{BlackHawk} to numerically compute both MW and EG contributions.
The differential flux of light DM from PBH evaporation in the MW is
\begin{equation}
\frac{d^2\phi_\chi^{\mathrm{MW}}}{dT_\chi\,d\Omega} = \frac{f_{\mathrm{PBH}}}{4\pi\,M_{\mathrm{PBH}}}\frac{d^2N_\chi}{dT_\chi\,dt}\int \frac{d\Omega_s}{4\pi}\int dl\, \rho_{\mathrm{MW}}[r(l,\psi)]\,,
\label{eq:mw_flux}
\end{equation}
where
$f_{\mathrm{PBH}}$ is the fraction of DM composed of PBHs, $\rho_{\mathrm{MW}}(r)$ is the DM density in the MW, and $r(l,\psi)=\sqrt{l^2+r_\odot^2-2lr_\odot \cos\psi}$ is the galactocentric distance with $r_\odot$ being the solar distance from the galactic center, $l$ the line-of-sight distance to the PBH and $\psi$ the angle between these two directions.
For $\rho_{\mathrm{MW}}(r)$, we adopt the Navarro-Frenk-White (NFW) profile~\cite{Navarro:1996gj,Wang:2020uvi}
\begin{equation}
\rho_{\mathrm{MW}}(r) = \rho_\odot \left(\frac{r}{r_\odot}\right)^{-\gamma}\left(\frac{1+r_\odot/r_s}{1+r/r_s}\right)^{3-\gamma}\,,
\label{eq:nfw_profile}
\end{equation}
where $\rho_\odot=0.4~\mathrm{GeV\,cm^{-3}}$ is the local density at the solar radius $r_\odot=8.5~\mathrm{kpc}$, $r_s=20~\mathrm{kpc}$ is the scale radius, and $\gamma=1$ characterizes the inner slope.
%
The flux from extragalactic PBHs is given by
\begin{equation}
\frac{d^2\phi_\chi^{\mathrm{EG}}}{dT_\chi\,d\Omega}
= \frac{n_{\mathrm{PBH}}(t_0)}{4\pi}
\int_{t_{\min}}^{t_0} \!dt\,[1+z(t)]\,
\left.\frac{d^2N_\chi}{dT_\chi\,dt}\right|_{E_\chi^s}\,.
\label{eq:eg_flux}
\end{equation}
where $z(t)$ is the redshift at time $t$, and the energy at the source is redshifted according to $E_\chi^s=\sqrt{(E_\chi^2-m_\chi^2)[1+z(t)]^2+m_\chi^2}$. For PBHs that can still constitute a fraction of DM today, we take their comoving number density as
\begin{equation}
n_{\mathrm{PBH}}(t_0)=f_{\mathrm{PBH}}\,\frac{\rho_{\mathrm{DM}}}{M_{\mathrm{PBH}}}\,,
\label{eq:npbh_comoving_f}
\end{equation}
with the mean cosmological DM density being $\rho_{\mathrm{DM}} = 2.35\times10^{-30}\,\mathrm{g\,cm^{-3}}$~\cite{Planck:2018vyg}.
Here $t_0$ is taken to be the current age of the Universe, which we take as $4.4 \times 10^{17}\,\mathrm{s}$.
In standard cosmology, matter--radiation equilibrium is reached at approximately $10^{11}$~s after the Big Bang. It is also worth noting that although we adopt this specific choice of $t_{\mathrm{min}}=10^{11}$ for the PBH formation time, we have explicitly verified that shifting it to an earlier time, even $t_{\mathrm{min}}=0$~s, has no significant effect on our results.

\begin{figure}[htbp]
\centering
\includegraphics[width=0.6\textwidth]{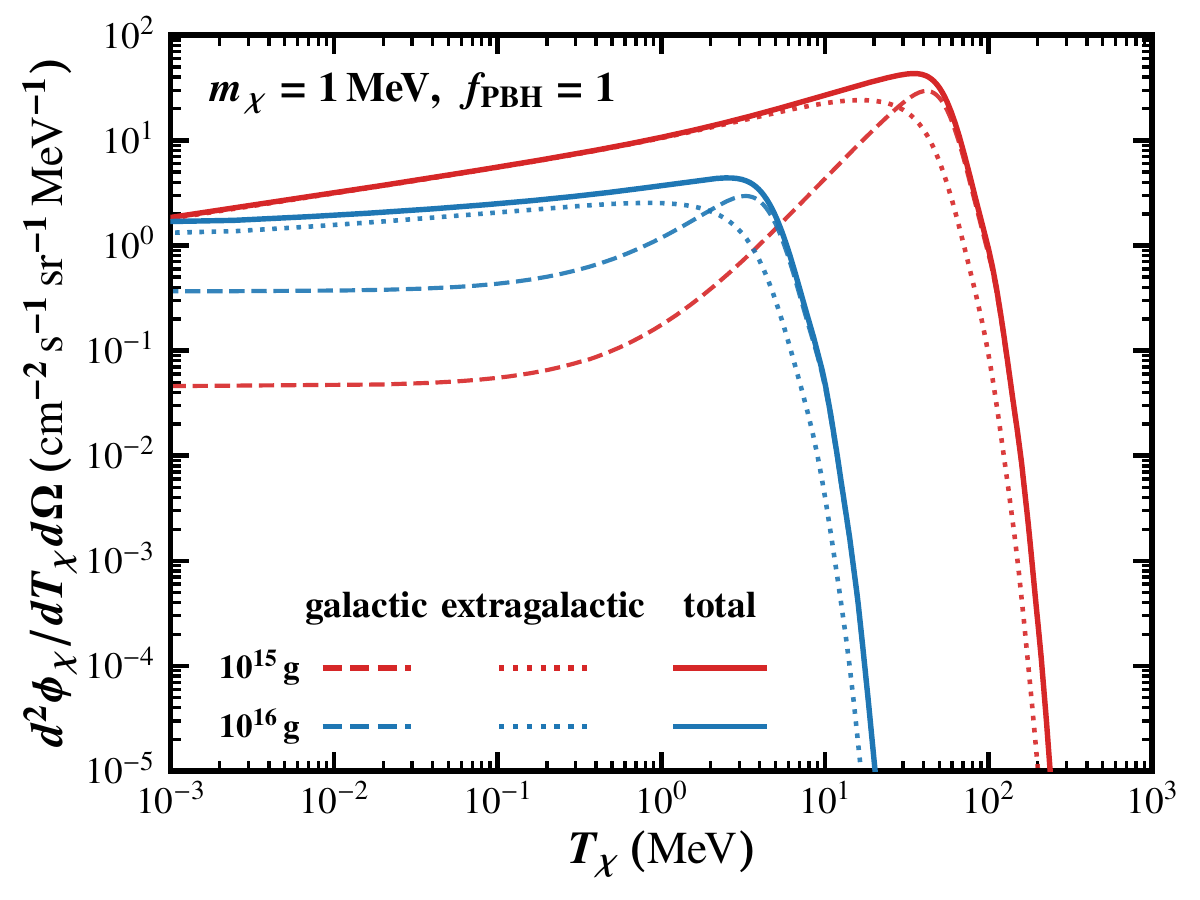}
\caption{
Differential DM flux at Earth as a function of the kinetic energy $T_\chi$ for $M_{\rm PBH}=10^{15}\,\mathrm{g}$ (red) or $10^{16}\,\mathrm{g}$ (blue), $m_\chi=1\,\mathrm{MeV}$ and $f_{\rm PBH}=1$. The dashed and dotted lines denote the galactic and extragalactic components, respectively, while solid lines show their sum.
}
\label{fig:PBHDMflux}
\end{figure}

The total differential flux is the sum of the galactic and extragalactic contributions, which is given by
\begin{equation}
\frac{d^2\phi_\chi}{dT_\chi\,d\Omega} = \frac{d^2\phi_\chi^{\mathrm{MW}}}{dT_\chi\,d\Omega} + \frac{d^2\phi_\chi^{\mathrm{EG}}}{dT_\chi\,d\Omega}\,.
\label{eq:flux_combined}
\end{equation}
In Fig.~\ref{fig:PBHDMflux}, we show the differential PBHBDM flux at Earth as a function of the kinetic energy $T_\chi$ for $f_{\rm PBH} = 1$ and $m_\chi = 1\,\mathrm{MeV}$. The dashed curves show the galactic contributions from PBHs in the MW and the dotted curves correspond to the extragalactic contribution. The solid curves denote the total flux given by the sum of the two components. We consider two representative PBH masses, $M_{\rm PBH}=10^{15}\,\mathrm{g}$ (red) and $M_{\rm PBH}=10^{16}\,\mathrm{g}$ (blue). Take $M_{\rm PBH}=10^{15}\,\mathrm{g}$ for illustration, one can see that the extragalactic contribution dominates the total flux for $T_\chi \lesssim 30$ MeV, whereas the MW contribution prevails at higher kinetic energies. Moreover, the flux peaks at $T_\chi\sim 40$ MeV which is much greater than $m_\chi$. For higher PBH mass, both the total flux and the cutoff energy decrease.

\subsection{Attenuation effect in the Earth matter}
\label{sec:attenuation}

As the DM particles produced by PBHs traverse the Earth, their flux will be attenuated due to their interaction with the SM particles in the terrestrial medium. Consequently, the DM flux arrived at an underground detector may significantly differ from the original one upon arrival on Earth due to the attenuation effect in the Earth matter. The energy loss per unit distance $x$ is expressed as~\cite{DeRomeri:2023ytt,Das:2024ghw}
\begin{equation}
\frac{dT_\chi}{dx} = - \ndens_i \int_{0}^{T_i^{\mathrm{max}}} dT_i\, T_i\, \frac{d\sigma_{\chi i}}{dT_i}\,,
\label{eq:attenuation_1}
\end{equation}
where $\ndens_i$ is the number density of electrons ($i=e$) with $\ndens_e=8\times10^{23}\,\mathrm{cm}^{-3}$~\cite{Ema:2018bih} or the number density of nuclei ($i=N$) with $\ndens_N=3.44\times10^{22}\,\mathrm{cm}^{-3}$~\cite{Kavanagh:2016pyr,Morgan:1980pnas,McDonough:2003treatise,Dziewonski:1981prem,DeRomeri:2023ytt}, $T_i$ is the kinetic energy transferred to the target, and $\frac{d\sigma_{\chi i}}{dT_i}$ is the differential cross section of $\chi$--$i$ scattering.
We assume that the differential cross section $\sigma_{\chi i}$ is independent of $T_i$ for the electron interactions and takes the form $\frac{d\sigma_{\chi i}}{dT_i}=\frac{\sigma_{\chi i}}{T_i^{\mathrm{max}}}$ with $T_i^{\mathrm{max}}$ being the maximum kinetic energy transferred to the target~\cite{DeRomeri:2023ytt}. For the nuclear interactions, the differential cross section is dependent on $T_i$ from the nuclear form factor, and we have used numerical method to do the integration of Eq.~(\ref{eq:attenuation_1}).

Under the assumptions of fully elastic scattering and appropriate relativistic corrections, the maximum kinetic energy transferable to the target is
\begin{equation}
T_i^{\mathrm{max}} = \frac{T_\chi^2 + 2m_\chi\,T_\chi}{T_\chi + \frac{(m_\chi+m_i)^2}{2m_i}}\,.
\label{eq:ti_max}
\end{equation}
where $m_i$ denotes the mass of electron or nucleus, and $m_\chi$ is the mass of light DM particles. By defining the inverse of the mean free path as
\begin{equation}
\ell^{-1} =  \ndens_i\,\sigma_{\chi i}\,\frac{2m_i m_\chi}{\left(m_i+m_\chi\right)^2}\,,
\label{eq:attenuation_6}
\end{equation}
one can solve the differential equation for $T_\chi$ and obtain the energy $T_{\chi}^d$ after the light DM propagates a distance $z$ to the detector~\cite{Das:2024ghw}
\begin{equation}
T_\chi^d = \frac{2m_\chi\,T_\chi^0\, e^{-z/\ell}}{T_\chi^0\left(1-e^{-z/\ell}\right)+2m_\chi}\,.
\label{eq:attenuation_7}
\end{equation}
The distance $z$ is given by~\cite{DeRomeri:2023ytt}
\begin{equation}
z = -(R_E - d)\cos\theta_z + \sqrt{R_E^2 - (R_E - d)^2\sin^2\theta_z}\,,
\label{eq:distance}
\end{equation}
where $R_E$ is the radius of the Earth, $\theta_z$ is the zenith angle of the detector, $d$ is the depth of the detector from the Earth surface at the zero zenith angle. $T_\chi^0$ is the initial kinetic energy at Earth surface. After inverting Eq.~(\ref{eq:attenuation_7}), we have~\cite{DeRomeri:2023ytt}
\begin{equation}
T_\chi^0 = \frac{2m_\chi\,T_\chi^d\, e^{z/\ell}}{2m_\chi+T_\chi^d\left(1-e^{z/\ell}\right)}\,.
\label{eq:attenuation_8}
\end{equation}
As a result, the attenuated DM flux becomes~\cite{Calabrese:2021src,Li:2022jxo}
\begin{equation}
\frac{d^2\phi_\chi^d}{dT_\chi^d\,d\Omega} = \frac{4m_\chi^2\,e^{z/\ell}}{\left[2m_\chi+T_\chi^d\left(1-e^{z/\ell}\right)\right]^2}\,\left.\frac{d^2\phi_\chi}{dT_\chi\,d\Omega}\,\right|_{T_{\chi}^0}\,.
\label{eq:attenuated}
\end{equation}

\begin{figure}[htbp]
\centering
\includegraphics[width=0.6\textwidth]{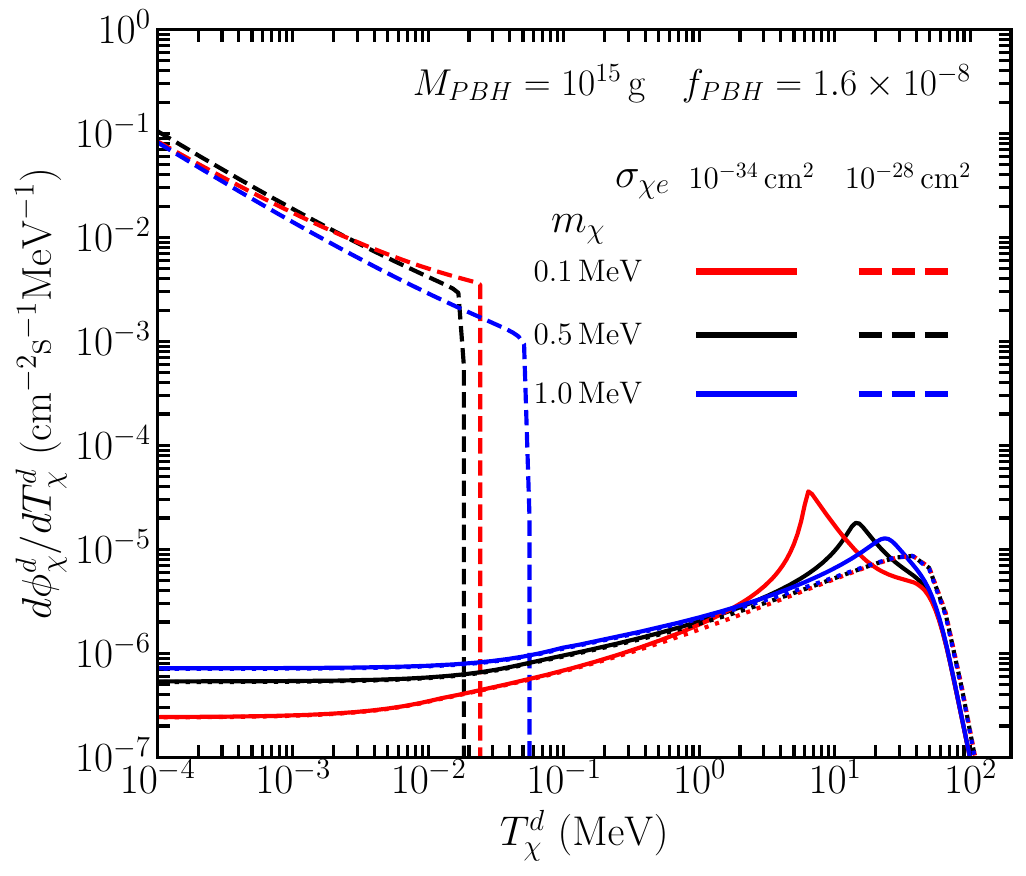}
\caption{
The PBHBDM flux distribution at the detector as a function of DM kinetic energy with $M_{\rm PBH}=10^{15}$ g, $f_{\rm PBH}=1.6\times 10^{-8}$ and the depth of detector $d=1$ km. We consider DM--electron scattering in Earth matter with ${\sigma_{\chi e}=10^{-28}\,\mathrm{cm^2}}$ (dashed) and ${\sigma_{\chi e}=10^{-34}\,\mathrm{cm^2}}$ (solid). The PBHBDM mass is taken as $m_{\chi}=0.1\,\mathrm{MeV}$ (red), $m_{\chi}=0.5\,\mathrm{MeV}$ (black) or $m_{\chi}=1.0\,\mathrm{MeV}$ (blue). The unattenuated flux reaching the detector is given by dotted line.
}
\label{fig:Attenuation}
\end{figure}

To demonstrate the dependence of attenuation effect on DM parameters, we take DM--electron scattering for illustration. Fig.~\ref{fig:Attenuation} shows the angle-integrated differential DM flux arriving at the detector $d\phi^d_\chi/dT^d_\chi$ as a function of DM kinetic energy. We assume ${\sigma_{\chi e}=10^{-28}\,\mathrm{cm^2}}$ (dashed) or ${\sigma_{\chi e}=10^{-34}\,\mathrm{cm^2}}$ (solid) for attenuation in Earth matter, and three DM masses: $m_\chi = 0.1\,\mathrm{MeV}$ (red), $0.5\,\mathrm{MeV}$ (black) and $1\,\mathrm{MeV}$ (blue). We also take $M_{\rm PBH}=10^{15}$ g and $f_{\rm PBH}=1.6\times 10^{-8}$. Note that here and below we choose this value for $f_{\rm PBH}$ as it is the maximally allowed value from the EDGES 21 cm constraint~\cite{Clark:2018ghm}. There exist weaker bounds from isotropic gamma-ray background (IGRB)~\cite{Carr:2009jm,Arbey:2019vqx,Chen:2021ngo} or other measurements~\cite{Boudaud:2018hqb,Coogan:2020tuf}. We also show the dotted curve corresponds to the unattenuated flux for comparison. One can see that as expected, the unattenuated flux peaks at several tens of MeV, and falls rapidly as $T_\chi^d$ increases.
For a small $\sigma_{\chi e}$ as shown by solid curves, the attenuation effect is relatively mild and the attenuated flux remains close to the unattenuated one. Also, through scattering in Earth matter, the attenuation effect shifts the flux at the Earth's surface to lower energies at the detector. As a result, the flux magnitude at low $T^d_\chi$ is enhanced and may exceed the unattenuated prediction. This pile-up phenomenon reaches its maximum at the maximum penetration depth $z$ with $\theta_z=\pi$.
Moreover, the attenuation is maximized when $m_\chi \simeq m_e$ because the energy transfer per collision is most efficient.
For the unattenuated spectrum peaks at $T_{\chi}^0 \simeq  40~\mathrm{MeV}$, from Eqs.~(\ref{eq:attenuation_7}) and (\ref{eq:distance}), we find the peak of attenuated flux with $\sigma_{\chi e} = 10^{-34}\,\mathrm{cm}^2$ and $m_{\chi}=0.5\,\mathrm{MeV}$ is shifted to $T_{\chi}^d \simeq  13~\mathrm{MeV}$, which agrees well with the solid black curve shown in Fig.~\ref{fig:Attenuation}. In contrast, the PBHBDM flux with a large $\sigma_{\chi e}$ given by dashed curve gains significantly strong attenuation.

\section{PBHBDM scattering in direct detection experiments}
\label{sec:scattering}

In DM DD experiments, DM scattering can transfer energy either to a target nucleus or to an electron. In this section, using the above attenuated flux of PBHBDM at detector, we will compute the expected event rate from PBHBDM scattering off electrons or nuclei in DM DD experiments.

\subsection{PBHBDM scattering off electron}

The event rate of DM--electron scattering in a terrestrial detector is given by~\cite{Li:2022jxo}
\begin{equation}
\frac{d^3N_\chi}{dtd\Omega\,dT_e} = \ntarget_e \int dT_\chi^d\, \sigma_{\chi e}\, D_\chi^e\left(T_e, T_\chi^d\right)\frac{d^2\phi_\chi^d}{dT_\chi^d\,d\Omega}\,,
\label{eq:basic_scattering}
\end{equation}
where $\ntarget_e$ is the number of electrons in the detector target, $\sigma_{\chi e}$ is the DM--electron scattering cross section, $T_e$ is the electron recoil (ER) energy. Here $\frac{d^{2}\phi_{\chi}^{\,d}}{dT_{\chi}^{\,d}\, d\Omega}$ denotes the attenuated DM flux at the detector from Eq.~(\ref{eq:attenuated}), in which we take $i=e$ for DM scattering off electrons.
The response function for ER scattering is
\begin{equation}
D_\chi^e\left(T_e, T_\chi^d\right) = \frac{1}{T_e^{\mathrm{max}}(T_\chi^d)}\,\Theta\Bigl(T_e^{\mathrm{max}}(T_\chi^d)-T_e\Bigr)\,,
\label{eq:response_function}
\end{equation}
where $T_e^{\mathrm{max}}(T_\chi^d)$ is the maximal recoil energy given the energy $T_\chi^d$ of a DM particle, and $\Theta$ is the Heaviside step function.

\begin{figure}[htbp]
\centering
\includegraphics[width=0.6\textwidth]{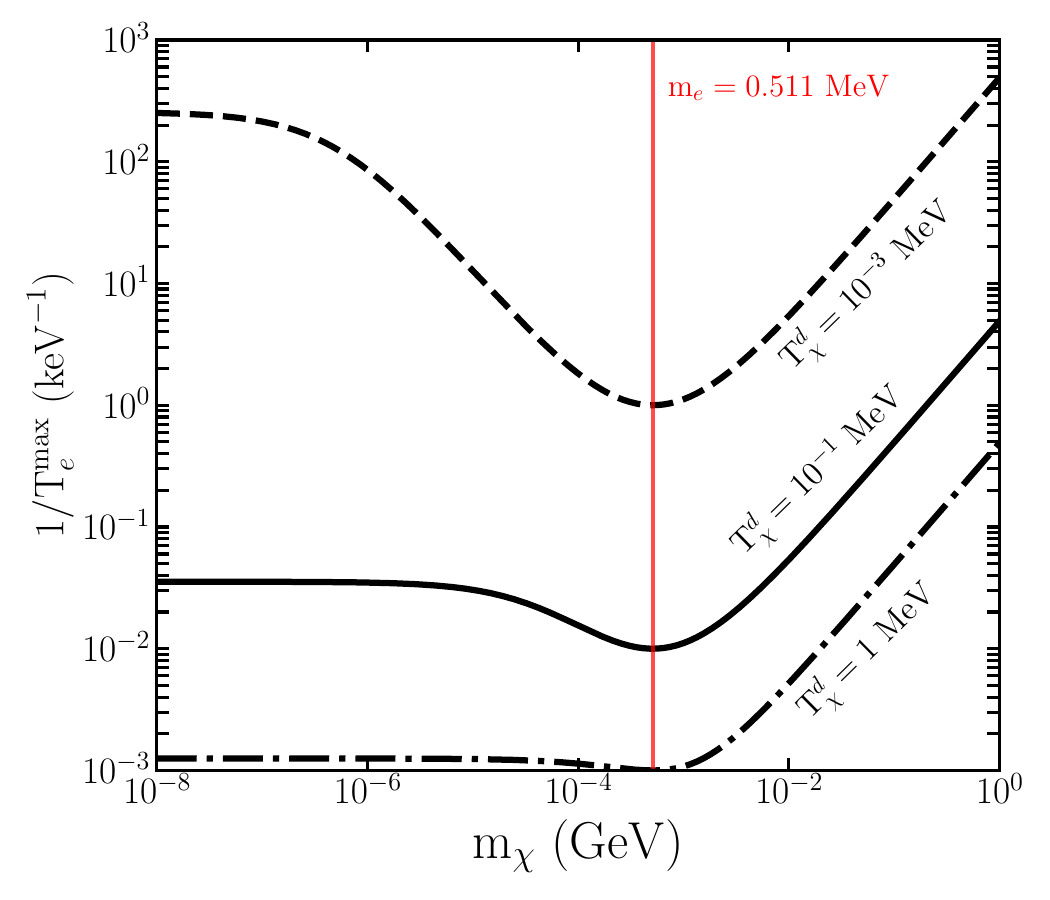}
\caption{$1/T_{e}^{\max}$ as a function of the DM mass $m_{\chi}$ at the detector with benchmark kinetic energies $T_{\chi}^{d} = 1\,\mathrm{MeV}$ (dash–dotted), $10^{-1}\,\mathrm{MeV}$ (solid) and $10^{-3}\,\mathrm{MeV}$ (dashed). The red solid vertical line indicates the mass of target particle $m_e$.
}
\label{fig:DIFF}
\end{figure}

We first examine how the response function in Eq.~(\ref{eq:response_function}) affects the relevant kinematic properties. We show $1/T_{e}^{\max}$ as a function of $m_\chi$ for different $T_{\chi}^d$ values in Fig.~\ref{fig:DIFF}.
Here, we consider four values of $T_{\chi}^d$, ranging from $10^{-3}\,\mathrm{MeV}$ to 1 MeV, where the highest energy is close to the electron mass. When inserted $T_{\chi}^d$ into the response function in Eq.~(\ref{eq:response_function}), the $\Theta$ function forces the observable recoil energy $T_e$ to lie below the corresponding $T_{e}^{\max}(T_\chi^d)$. Thus, only the region with $T_e<T_{e}^{\max}(T_\chi^d)$ contributes to the integral in Eq.~(\ref{eq:basic_scattering}). As shown in Fig.~\ref{fig:DIFF}, $D_{\chi}^e$ exhibits a dip for $m_\chi$ near electron mass and then increases beyond the electron mass. For small $\sigma_{\chi e}$, the response function described by Eq.~(\ref{eq:response_function}) dominates the event rates over the negligible attenuation effect. This variation of $D_{\chi}^e$ thus causes the expected event rates to first decrease near $m_e$ and then arise as $m_\chi$ increases.

Given the above differential flux, we can calculate the predicted event rate for each DD experiment as follows
\begin{equation}
\frac{dN_{\chi}}{dT_{e}}
=
\ntarget_{\mathrm{Xe}}\, t_{\mathrm{expo}}
\int d\Omega\, dT_{\chi}^{\,d}
Z_{\mathrm{eff}}^{\mathrm{Xe}}(T_e)\,
\sigma_{\chi e}\,
D_\chi^e\left(T_e, T_\chi^d\right)\,
\frac{d^{2}\phi_{\chi}^{\,d}}{dT_{\chi}^{\,d}\, d\Omega}\,,
\label{electron_scattering}
\end{equation}
where $\ntarget_{\mathrm{Xe}}$ and $t_{\mathrm{expo}}$ denote the target number of xenon in the detector and the exposure time, respectively. $Z_{\mathrm{eff}}^{\mathrm{Xe}}(T_e)$ is xenon's effective electron charge as a function of recoil energy $T_e$~\cite{AtzoriCorona:2022jeb} and takes the form $Z_{\mathrm{eff}}^{\mathrm{Xe}}(T_e)=\sum_i^{Z^{\mathrm{Xe}}}\Theta(T_e-B_i)$ with $B_i$ being the binding energy of the $i$th electron~\cite{Chen:2016eab}. It accounts for the binding effect of atomic electrons in xenon and  quantifies the number of electrons ionized by a certain energy deposit $T_e$.
After taking into account the energy resolution, we have
\begin{equation}
\frac{dN_{\chi}}{dT_{e}^{\rm reco}}
=
F_e(T_{e}^{\rm reco})
\int_{0}^{T_e^{\rm max}} dT_e\;
G_{e}\!\left(T_{e}^{\rm reco}, T_{e}\right)\,
\frac{dN_{\chi}}{dT_{e}}\,,
\label{eq:rec_events_er}
\end{equation}
where $T_{e}$ ($T_{e}^{\rm reco}$) denotes the true (reconstructed) energy, $F_e$ represents the detection efficiency provided by each experiment. We take the approach in Ref.~\cite{DeRomeri:2023ytt} for the efficiency and detector smearing. For the DM--electron scattering calculation, the XENONnT~\cite{XENON:2022ltv} and PandaX~\cite{PandaX:2024cic} experiments employ the reconstructed energy efficiency $F_e(T_e^{\mathrm{reco}})$, whereas the LZ~\cite{LZ:2022lsv} utilizes the true recoil energy efficiency $F_e(T_e)$. $G_{e}\!\left(T_{e}^{\rm reco}, T_{e}\right)$ is the following gaussian smearing function used to simulate detector resolution that maps $T_{e}$ to $T_{e}^{\rm reco}$, i.e.,
\begin{equation}
G_{e}\!\left(T_{e}^{\rm reco}, T_{e}\right)
=\frac{1}{\sqrt{2\pi}\,\sigma_{e}(T_{e})}
\exp\!\left[-\frac{\big(T_{e}^{\rm reco}-T_{e}\big)^{2}}{2\,\sigma_{e}^{2}(T_{e})}\right]\,,
\label{gauss_defined}
\end{equation}
where $\sigma_{e}(T_{e})$ is the energy resolution in unit of keV. We take $\sigma_{\mathrm{e}}(T_e)=0.31\sqrt{T_e}+3.7\times10^{-3}\,T_e$ for XENONnT~\cite{XENON:2020rca}, $\sigma_{\mathrm{e}}(T_e)=0.073+0.173\,T_e-6.5\times10^{-3}T_e^{2}+1.1\times10^{-4}T_e^{3}$ for PandaX~\cite{PandaX:2022ood}, and $\sigma_{\mathrm{e}}(T_e)=0.323\sqrt{T_e}$ for LZ~\cite{A:2022acy}.

\begin{figure}[htbp]
\centering
\includegraphics[width=0.46\textwidth]{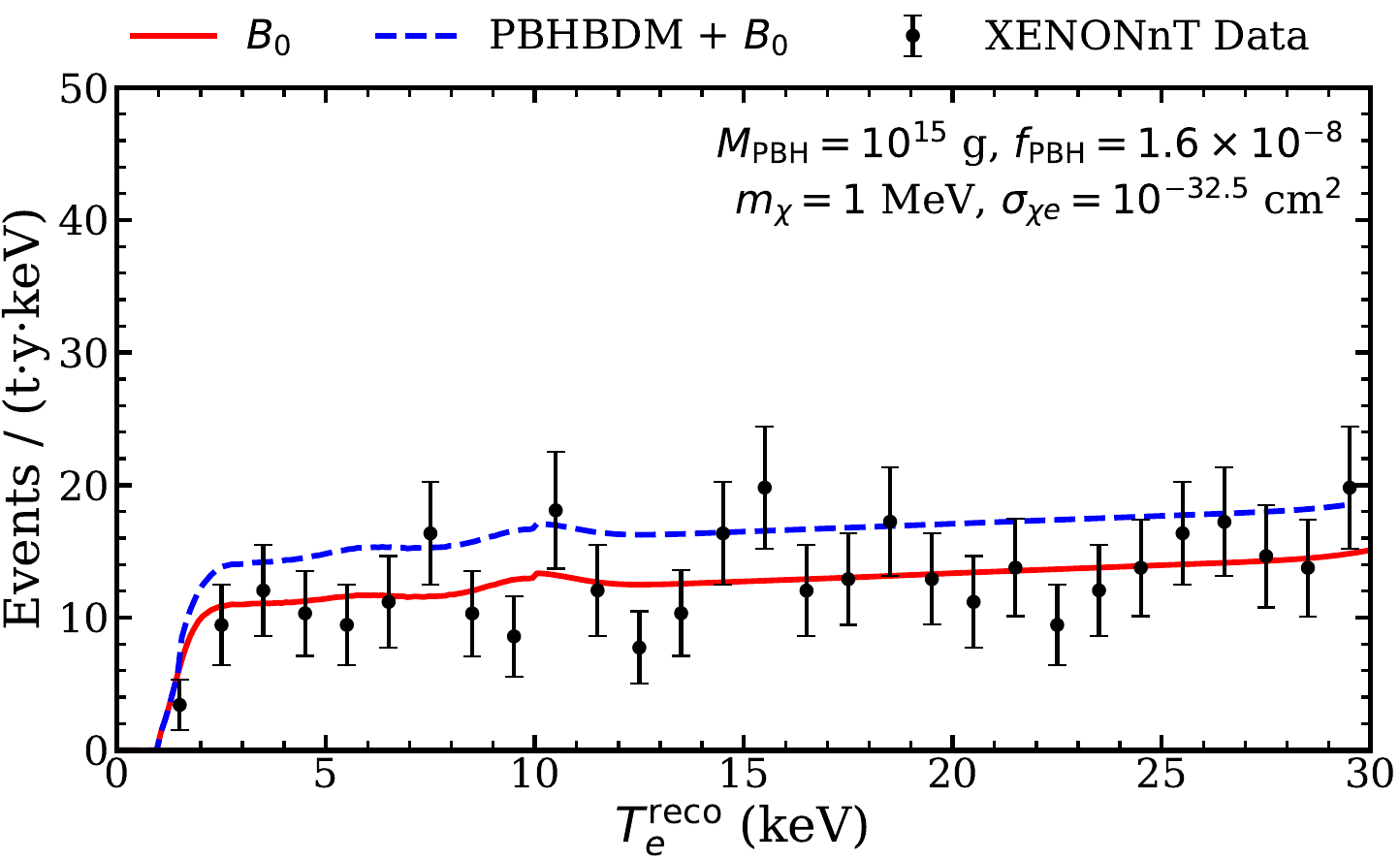}\hfill
\includegraphics[width=0.46\textwidth]{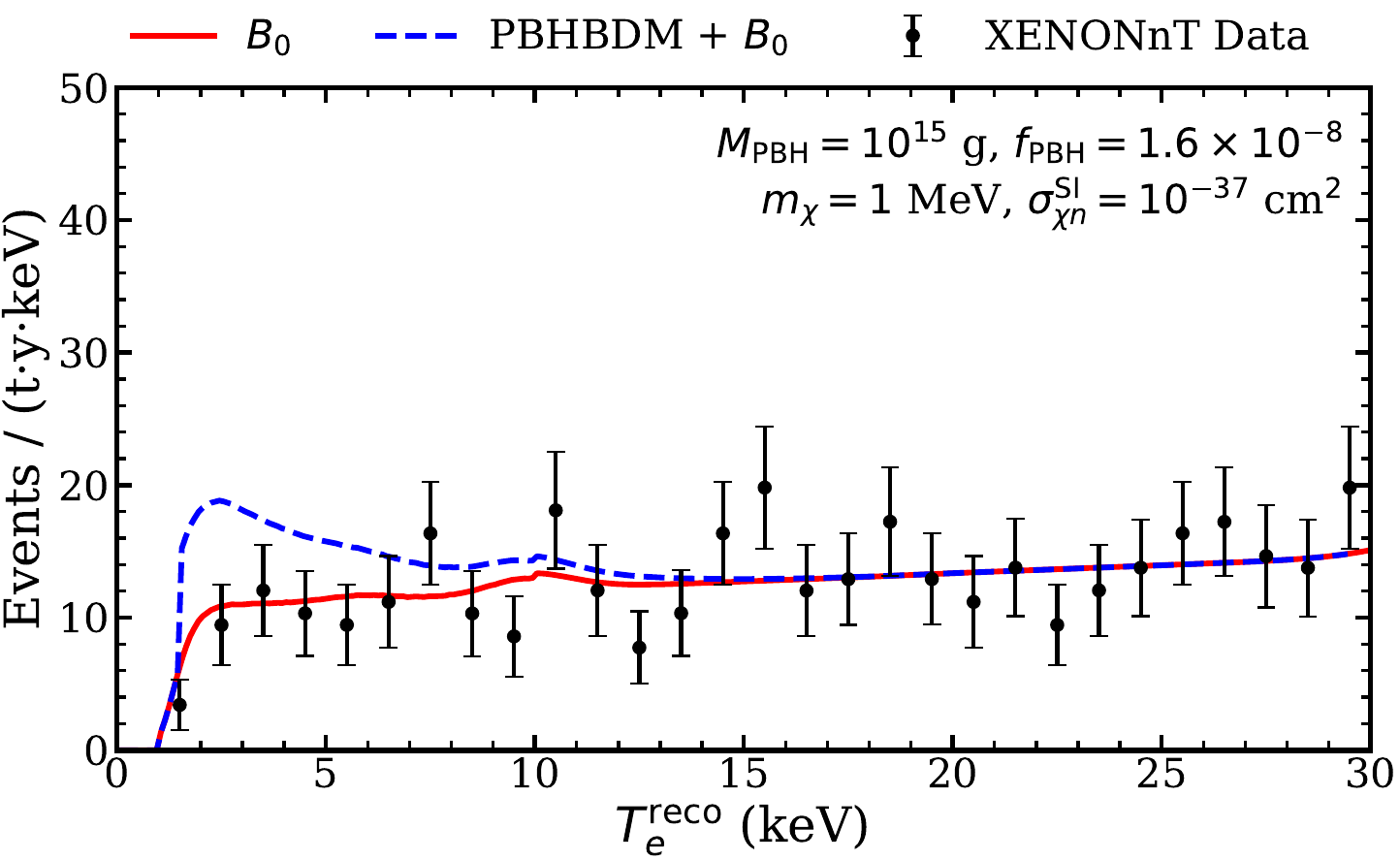}\hfill
\includegraphics[width=0.46\textwidth]{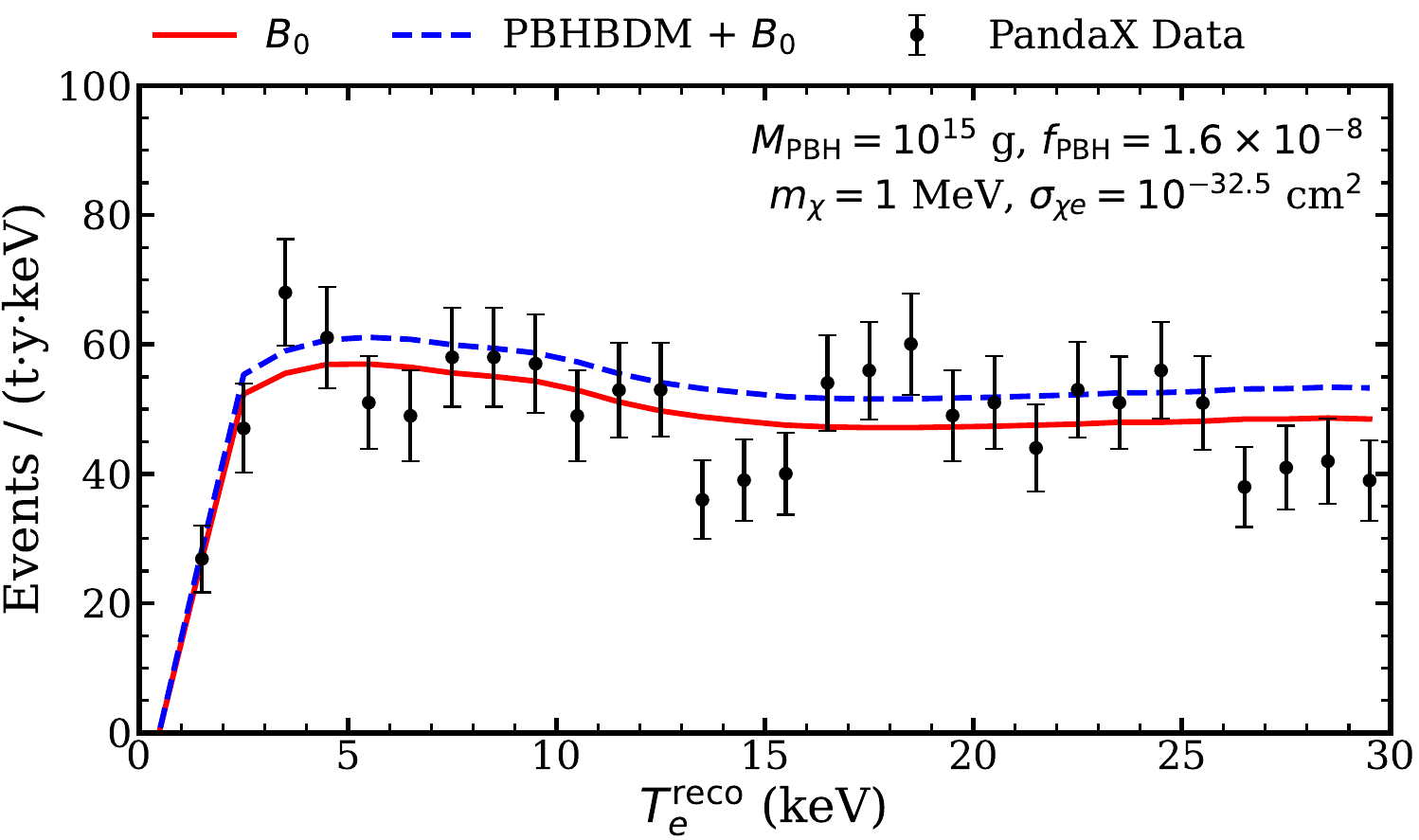}\hfill
\includegraphics[width=0.46\textwidth]{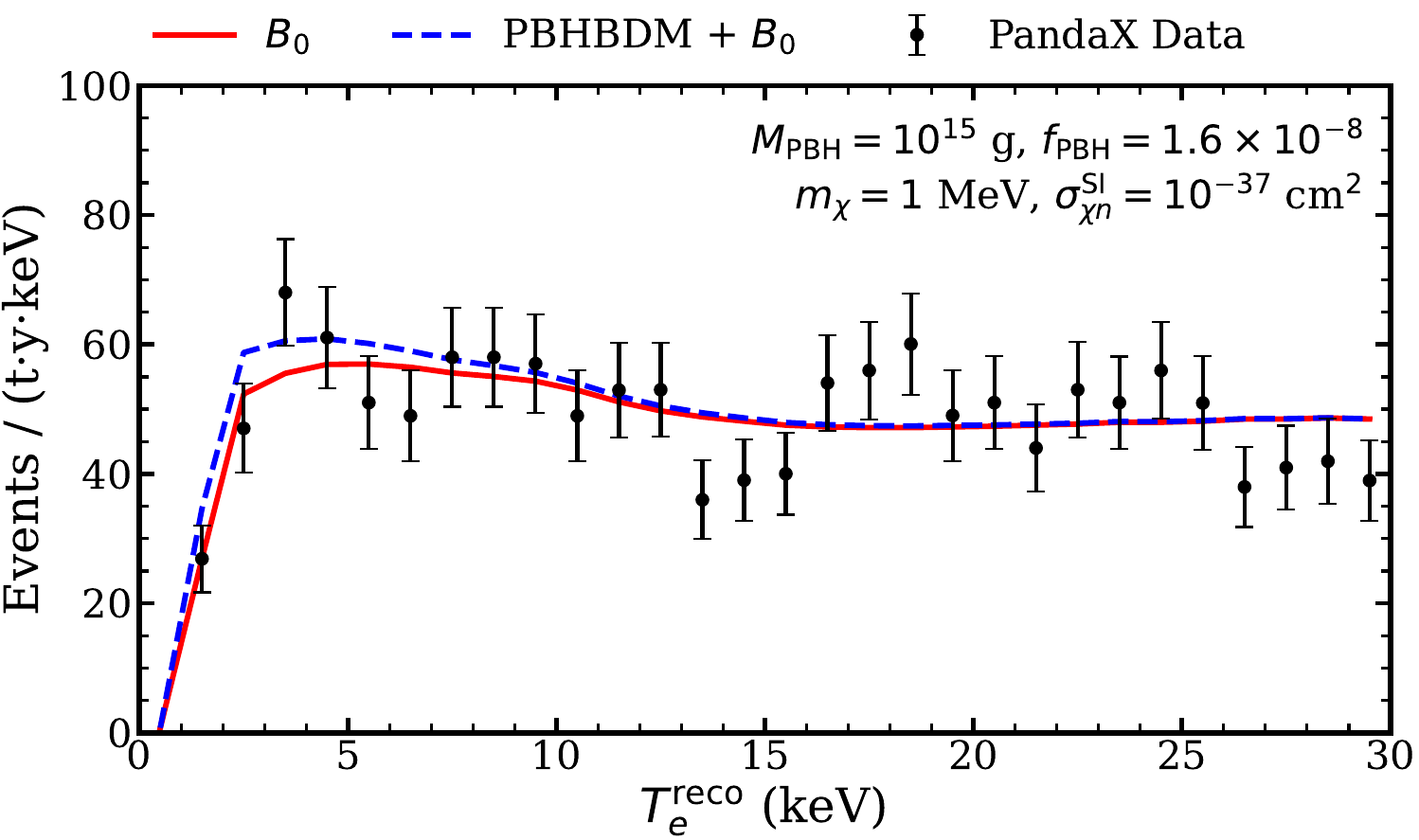}\hfill
\includegraphics[width=0.46\textwidth]{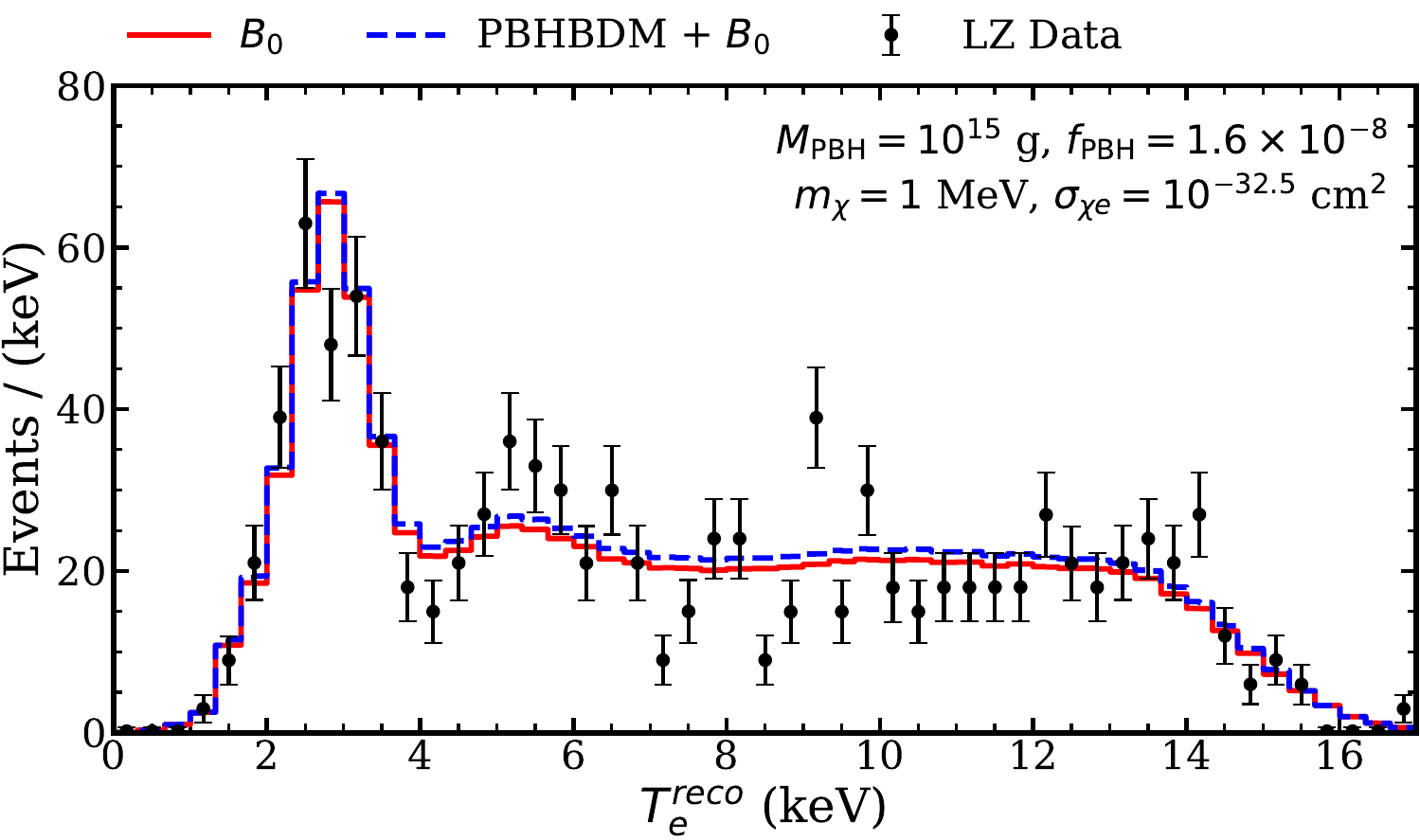}\hfill
\includegraphics[width=0.46\textwidth]{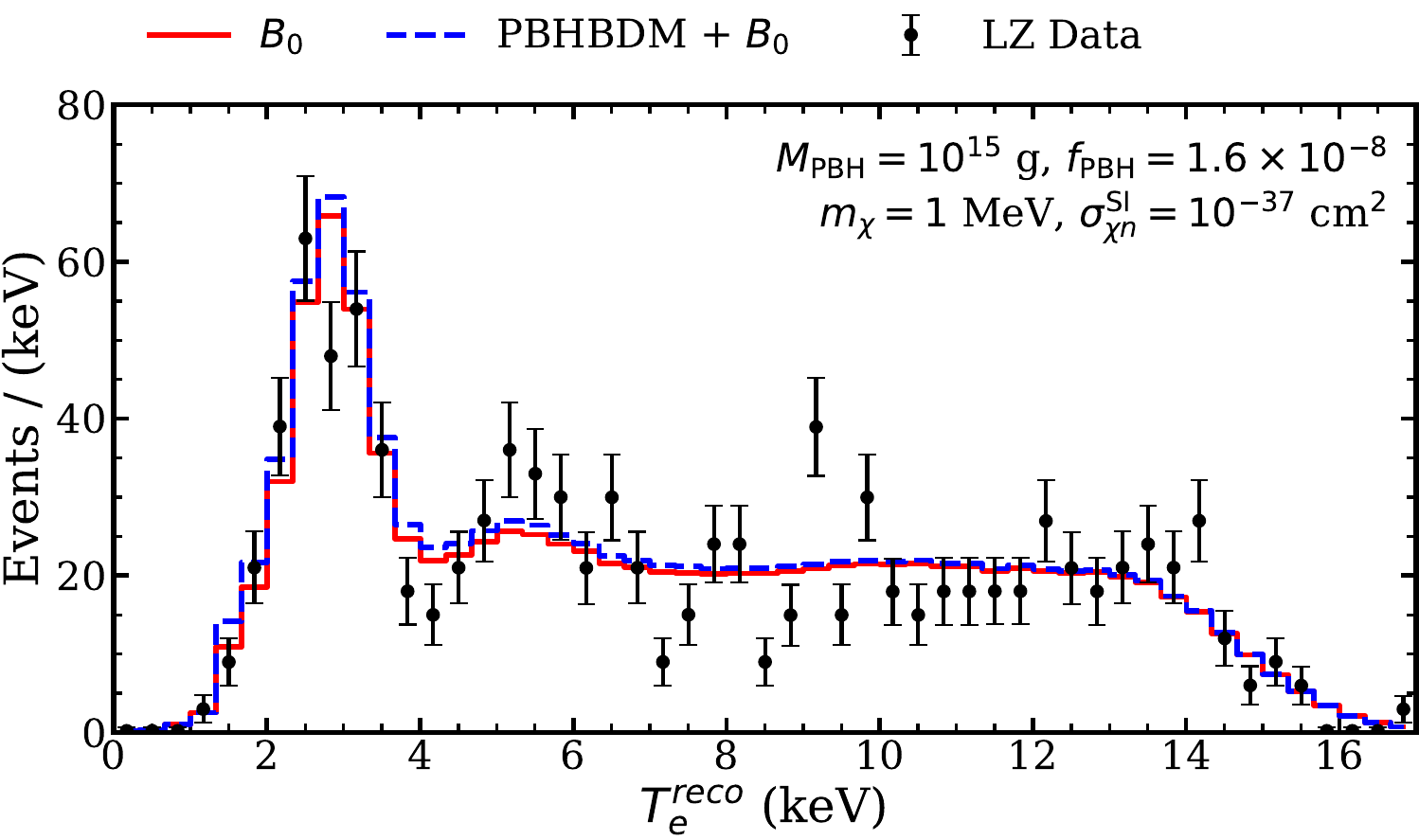}
\caption{The predicted event spectrum of DM-electron scattering (left) and DM--nucleus scattering (right) for XENONnT (top), PandaX (middle), and LZ (bottom) experiments. Black points denote the measured data with error bars, and the red solid curve is the background-only spectrum provided by each experiment. The blue dashed curve shows the total expectation event spectrum including the DM signal induced by PBHs also. The benchmark parameters are $M_{\rm PBH}=10^{15}\,\mathrm{g}$, $f_{\rm PBH}=1.6\times10^{-8}$, $m_\chi=1\,\mathrm{MeV}$, $\sigma_{\chi e}=10^{-32.5}\,\mathrm{cm}^2$ and \(\sigma_{\chi n}^{\rm SI}=10^{-37}\,\mathrm{cm^2}\).
}
\label{fig:ER_NR_events}
\end{figure}

We use Eq.~(\ref{eq:rec_events_er}) to calculate the expected number of events from DM--electron scattering. The left panels of Fig.~\ref{fig:ER_NR_events} shows the reconstructed electron recoil spectra as functions of the reconstructed energy $T_e^{reco}$ for XENONnT (top), PandaX (middle), and LZ (bottom) experiments. The black points with error bars denote the measured data reported by each experiment, while the red solid curves show the background-only expectation $B_0$. The blue dashed curves represent the total prediction together with the signal events arising from PBHBDM--electron scattering. The benchmark parameters used to generate the signal spectra include $M_{\rm PBH}=10^{15}\,{\rm g}$, $f_{\rm PBH}=1.6\times10^{-8}$, $m_\chi=1\,{\rm MeV}$, and $\sigma_{\chi e}=10^{-32.5}\,{\rm cm}^2$.
One can seen that XENONnT and PandaX experiments exhibit good agreement between the data and background. As a result, we expect that they would induce strong constraints on the parameters of PBHBDM scenario.

\subsection{PBHBDM scattering off nucleus}

For the case of DM--nucleus scattering, the recoil energy is primarily deposited through elastic collisions with nuclei. Most of the recoil energy is converted into heat and atomic motion rather than ionization. The event rate induced by DM--nucleus scattering at a DM DD detector is given by
\begin{equation}
\frac{dN_{\chi}}{dT_{N}}
= \ntarget_{\mathrm{Xe}}\, t_{\mathrm{expo}}\,F_N(T_N)\int d\Omega\, dT_{\chi}^{\,d}\;
\sigma_{\chi N}^{\rm SI}\,
D_{\chi}^{N}\!\left(T_{N},\,T_{\chi}^{\,d}\right)\,
\frac{d^{2}\phi_{\chi}^{\,d}}{dT_{\chi}^{\,d}\, d\Omega}\,,
\label{eq:nucleus_scattering}
\end{equation}
where $T_{N}$ is the true nuclear recoil (NR) energy while $F_N(T_N)$ is defined as the true recoil efficiency that applies at the true energy level before smearing to $T_e^{\mathrm{reco}}$. Regarding DM--nucleus scattering, our calculations incorporate the efficiencies provided by XENONnT~\cite{XENON:2023cxc}, PandaX~\cite{PandaX-4T:2021bab} and LZ~\cite{LZ:2023poo}. $\sigma_{\chi N}^{\rm SI}$ is the SI DM--nucleus scattering cross section. The response function $D_{\chi}^{N}$ is obtained from Eq.~(\ref{eq:response_function}) by replacing electron quantities with nuclear counterparts, and $\frac{d^{2}\phi_{\chi}^{\,d}}{dT_{\chi}^{\,d}\, d\Omega}$ is the attenuated DM flux for DM--nucleus scattering obtained from Eq.~(\ref{eq:attenuated}).
The SI DM--nucleus cross section at momentum transfer \(q\) is~\cite{Calabrese:2021src}
\begin{equation}
  \sigma_{\chi N}^{\rm SI}(q^2)
  = \frac{\mu_N^2}{\mu_n^2}\,A^2\,\sigma_{\chi n}^{\rm SI}\,F^2(q^2)\,,
  \label{eq:si_chiN_q2}
\end{equation}
where \(\mu_N = \frac{m_\chi m_N}{m_\chi + m_N}\), and \(\mu_n = \frac{m_\chi m_n}{m_\chi + m_n}\), \(A\) is the nuclear mass number. For the attenuation calculation we adopt an effective average value \(A=33.3\)~\cite{DeRomeri:2023ytt}, whereas for LXe targets we take \(A=131\). Here \(m_n\) is the nucleon mass with \(m_N\simeq A\,m_n\), \(\sigma_{\chi n}^{\rm SI}\) is the SI DM--nucleon cross section, and \(F(q^2)\) is the nuclear form factor, which we have taken as the Helm form factor~\cite{Lewin:1995rx}. To incorporate the energy-dependent form factor, we follow the procedure given in Refs.~\cite{Das:2024ghw,Jeesun:2026wcx} and employ a numerical method to evaluate the attenuation effect during the propagation in the Earth and the nucleus scattering process in the detector.

Since the XENONnT, PandaX, and LZ collaborations have all reported their measured results in terms of electron--recoil signals, we follow Ref.~\cite{DeRomeri:2023ytt} and use the quenching factor to convert the nuclear recoil energies into ``electron-equivalent'' recoil energies. The quenching factor $Q_f(T_N)\equiv T_e/T_N$ provides a theoretical method to describe the suppression of scintillation and ionization signals associated with nuclear recoils. Here we take the standard Lindhard quenching factor~\cite{Barker:2012ek}
\begin{equation}
Q_f(T_N)=\frac{k\,g(\epsilon)}{1+k\,g(\epsilon)}\,,
\label{eq:lindhard_quenching}
\end{equation}
where \(g(\epsilon)=3\epsilon^{0.15}+0.7\epsilon^{0.6}+\epsilon\), \(\epsilon = 11.5\,Z^{-7/3}\left(T_N/\mathrm{keV}\right)\), and \(k = 0.133\,Z^{2/3}A^{-1/2}\) with $Z$ being atomic number. For a LXe target we take \(Z=54\) and \(A\simeq131\), which gives \(k\simeq 0.166\).

Once the differential rate of nuclear recoil events $dN_{\chi}/dT_N$ is computed, the equivalent electron--recoil rate $dN_{\chi}/dT_e$ can be obtained via the following transformation~\cite{DeRomeri:2023ytt}
\begin{equation}
  \frac{dN_{\chi}}{dT_e}
  =
  \frac{\displaystyle \frac{dN_{\chi}}{dT_N}}
       {Q_f(T_N)
        +
        T_N\,\frac{dQ_f}{dT_N}}
  \,,
  \label{eq:NR2ER_conversion}
\end{equation}
where the derivative term $dQ_f/dT_N$ accounts for the energy variation of the quenching factor. The energy-dependent quenching factor $Q_f(T_N)$ shifts events toward lower electron-equivalent energies.
Following Eq.~(\ref{eq:NR2ER_conversion}), the differential number of events from DM--nucleus scattering is given by
\begin{equation}
\frac{dN_{\chi}}{dT_{e}^{\rm reco}}
=\,\int_0^{T_e^{max}} dT_e\,G_{e}\!\left(T_{e}^{\rm reco}, T_{e}\right)\,\frac{dN_{\chi}}{dT_{e}}\,.
\label{events of NR}
\end{equation}
The right panels in Fig.~\ref{fig:ER_NR_events} show the predicted electron-equivalent NR spectra for the three experiments. For the DM--nucleus scattering signal we adopt cross section $\sigma_{\chi n}^{\rm SI}=10^{-37}\,{\rm cm}^2$, and fix $M_{\rm PBH}=10^{15}\,{\rm g}$, $f_{\rm PBH}=1.6\times10^{-8}$, and $m_\chi=1\,{\rm MeV}$. As we see from the right panels of Fig.~\ref{fig:ER_NR_events}, the signal event rates in the low-energy region for all three experiments, but only the event spectrum of XENONnT shows a significant distinction between total prediction and experimental data for the chosen benchmark parameters.

\section{Constraints on the light DM and PBH parameter space}
\label{sec:results}

For the analysis of XENONnT and PandaX data, we adopt the following Gaussian $\chi^{2}$ function~\cite{Almeida:1999ie,DeRomeri:2023ytt}
\begin{equation}
\chi^{2}(\vec S)
=\sum_{i=1}^{30}\left[
\frac{N^{\,i}_{\mathrm{pred}}(\vec S)-N^{\,i}_{\mathrm{exp}}}{\sigma^{\,i}}
\right]^{2}\,,
\label{eq:chis_gaussian}
\end{equation}
where \(N^{\,i}_{\mathrm{pred}}(\vec S)=N^{\,i}_{\mathrm{PBHBDM}}(\vec S)+B^{\,i}_{0}\) is the predicted event counts at the $i$th bin, and \(B_0^i\) (\(N^{i}_{\mathrm{exp}}\)) denotes the experimental background (data) extracted from XENONnT~\cite{XENON:2022ltv} and PandaX~\cite{PandaX:2024cic}. Here $\vec S$ includes the new physics parameters (e.g., $m_\chi$ and $f_{\mathrm{PBH}}$), and \(\sigma^{i}\) is the experimental uncertainty at the $i$th bin.

For the LZ analysis, we employ the following Poissonian $\chi^{2}$ function due to the low counts of LZ data~\cite{DeRomeri:2023ytt}
\begin{equation}
\chi^{2}(\vec S;\alpha,\delta)
=2\sum_{i=1}^{51}\!\left[
 N^{\,i}_{\mathrm{pred}}(\vec S;\alpha,\delta)
 - N^{\,i}_{\mathrm{exp}}
 + N^{\,i}_{\mathrm{exp}}\ln\!\frac{N^{\,i}_{\mathrm{exp}}}{N^{\,i}_{\mathrm{pred}}(\vec S;\alpha,\delta)}
\right]
+\left(\frac{\alpha}{\sigma_{\alpha}}\right)^{2}
+\left(\frac{\delta}{\sigma_{\delta}}\right)^{2}\,,
\label{eq:chis_poisson}
\end{equation}
with $N^{\,i}_{\mathrm{pred}}(\vec S;\alpha,\delta)=(1+\alpha)\,N^{\,i}_{\mathrm{bkg}}+N^{\,i}_{\mathrm{PBHBDM}}(\vec S)+(1+\delta)\,N^{\,i}_{{}^{37}\mathrm{Ar}}$.
Here \(\alpha,\delta\) are nuisance priors for the non-\({}^{37}\)Ar background and the \({}^{37}\)Ar background, respectively~\cite{AtzoriCorona:2022jeb}. We take \(\sigma_{\alpha}=13\%\)~\cite{Horiuchi:2008jz} and \(\sigma_{\delta}=100\%\)~\cite{DeRomeri:2023ytt}. In our analysis, we use \(\Delta\chi^{2}\equiv\chi^{2}-\chi^{2}_{0}=6.18\) to get the \(2\sigma\) exclusion regions in the parameter space of \((m_{\chi},\,\sigma_{\chi e})\) and \((m_{\chi},\,\sigma_{\chi n}^{\rm SI})\), and \(\Delta\chi^{2}\equiv\chi^{2}-\chi^{2}_{0}=4.0\) for the limits in \((f_{\mathrm{PBH}},\,M_{\mathrm{PBH}})\) parameter space.
Here $\chi^{2}_{0}$ corresponds to the background-only hypothesis, and we find $\chi^{2}_{0}=20.7, 29.0$ and $94.0$ for XENONnT, PandaX, and LZ, respectively.

\subsection{Exclusion regions in the light DM parameter space}

\begin{figure}[!t]
\centering
\includegraphics[width=0.5\textwidth]{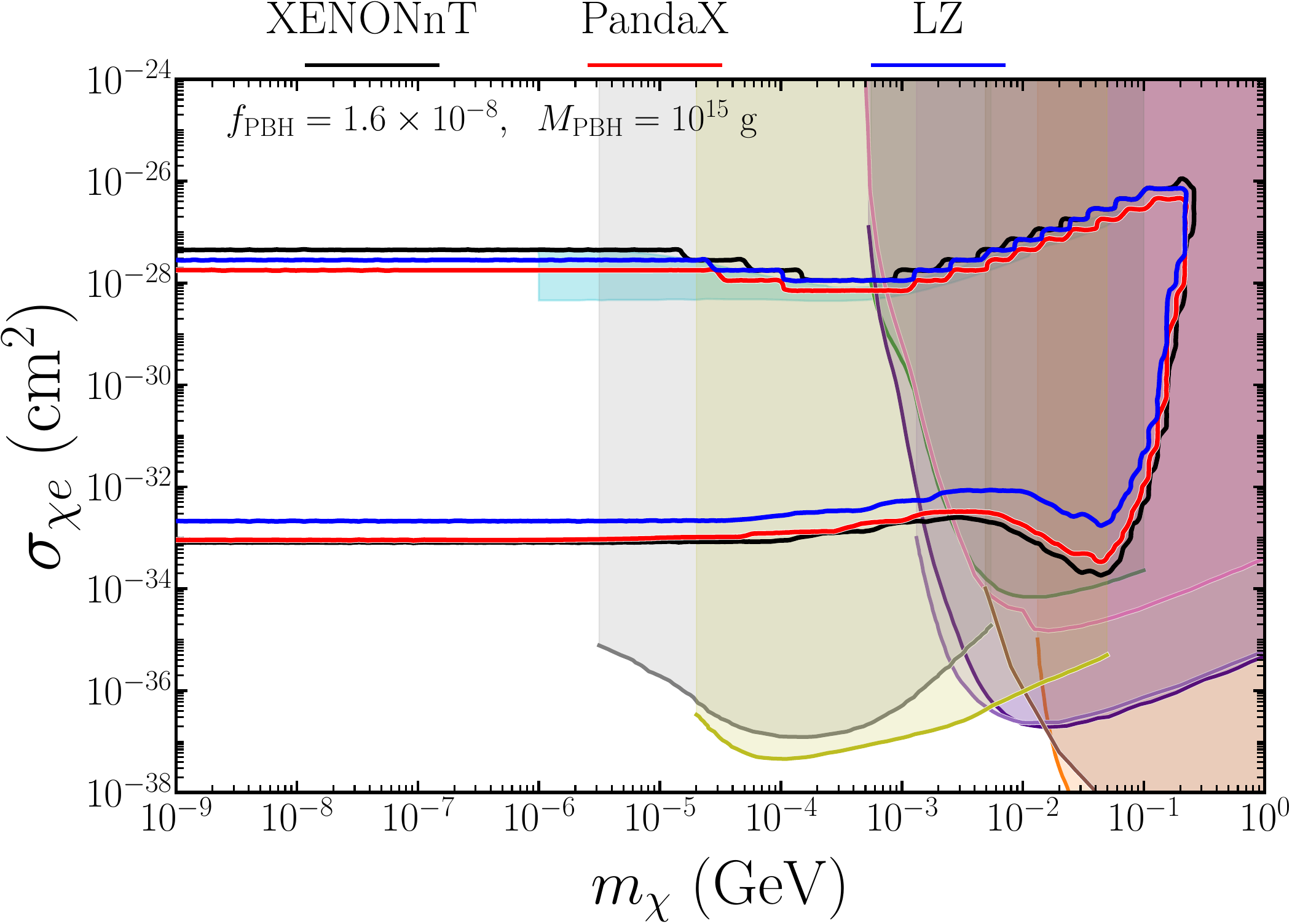}\hfill
\includegraphics[width=0.5\textwidth]{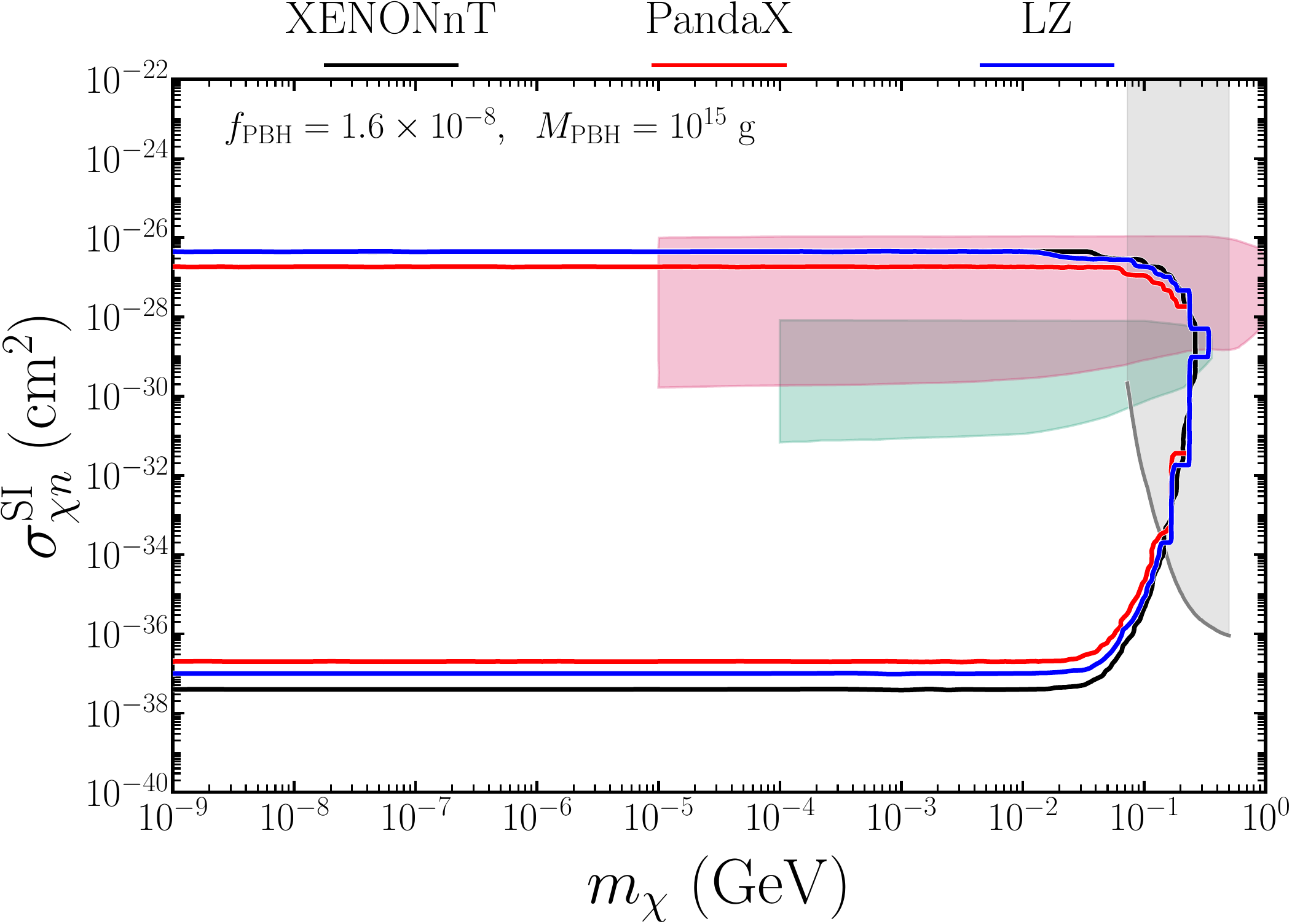}
\caption{Left: \(2\sigma\) bounds on \(\sigma_{\chi e}\) as a function of \(m_{\chi}\). Other constraints from DarkSide-50~\cite{DarkSide:2022knj}~(orange), EDELWEISS2020~\cite{EDELWEISS:2020fxc}~(green), DAMIC-M~\cite{DAMIC-M:2023gxo}~(indigo), SENSEI2025~\cite{SENSEI:2023zdf}~(purple), SuperCDMS2025~\cite{SuperCDMS:2024yiv}~(pink), and XENONnT~\cite{XENON:2024znc}~(brown) are shown for comparison. There exist more model-dependent constraints in the keV--MeV mass range from transmons~\cite{Hochberg:2026mdl} and nanowires~\cite{QROCODILE:2024nqm}. We also show the $90\%$ CL exclusion regions from CDEX-10~\cite{CDEX:2024xqm}~(cyan) for PBHBDM with the same $f_{\rm PBH}$ and $M_{\rm PBH}$ parameters. The earlier solar-reflection limit~\cite{An:2017ojc}~(gray) and the updated solar-reflection result~\cite{An:2021qdl}~(olive) are also shown for reference.
Right: \(2\sigma\) bounds on \(\sigma_{\chi n}^{\mathrm{SI}}\) as a function of \(m_{\chi}\) for Helm form factor. The gray region shows the constraint from {CRESST}~\cite{CRESST:2022lqw}~(gray). The cosmic-ray boosted DM exclusion regions from PandaX~\cite{PandaX-II:2021kai}~(teal) and CDEX~\cite{CDEX:2022fig}~(magenta) are also shown for comparison.
Both panels assume \(M_{\mathrm{PBH}}=10^{15}\,\mathrm{g}\) and \(f_{\mathrm{PBH}}=1.6\times10^{-8}\). The bounds are from XENONnT (black line), PandaX (red line) and LZ (blue line).
}
\label{fig:mchi_sigmae_pair}
\end{figure}

The regions excluded by the XENONnT, PandaX and LZ experiments for both DM--electron and DM--nucleus scattering are shown in the left and right panels of Fig.~\ref{fig:mchi_sigmae_pair} respectively, with \(M_{\mathrm{PBH}}=10^{15}\,\mathrm{g}\) and \(f_{\mathrm{PBH}}=1.6\times10^{-8}\). The left panel shows the 2\(\sigma\) exclusion regions in the plane of $\sigma_{\chi e}$ vs. $m_\chi$.
We can see that the upper bound exhibits a significant dip at $\sigma_{\chi e} \simeq 1\times 10^{-28}\,\mathrm{cm}^2$ and $m_\chi\simeq 0.5$ MeV. As seen from Fig.~\ref{fig:Attenuation} and previous discussion, a larger $\sigma_{\chi e}$ results in a greater attenuation effect, and the strongest attenuation effect occurs at $m_\chi\simeq m_e$. Therefore, a lower cross section is needed here to achieve the same attenuation result as other $m_\chi$.
The lower bounds also become weaker around $\sigma_{\chi e} \simeq 2\times 10^{-34}\,\mathrm{cm}^2$ and $m_\chi\simeq m_e$. As seen in Eq.~(\ref{eq:response_function}) and Fig.~\ref{fig:DIFF}, for the low $\sigma_{\chi e}$ regime, the event rate is strongly suppressed at $m_{\chi} \simeq m_e$. This is the region where a distinct dip is present in Fig.~\ref{fig:DIFF} and the bump of $\sigma_{\chi e}$ happens in Fig.~\ref{fig:mchi_sigmae_pair}. Note that PBHs with masses around $10^{15}~\mathrm{g}$ are relatively stable and have not yet reached the final explosive phase. They emit particles with a constant Hawking temperature. Thus, the Boltzmann factor suppresses the production of heavier DM and leads to a cutoff at the right-hand side of the excluded region.
We find the region of \(7.8\times10^{-34}~\mathrm{cm}^2 \lesssim \sigma_{\chi e} \lesssim 4.5\times10^{-28}~\mathrm{cm}^2\) (\(9\times10^{-34}~\mathrm{cm}^2 \lesssim \sigma_{\chi e} \lesssim 1.9\times10^{-28}~\mathrm{cm}^2\)) [\(2\times10^{-33}~\mathrm{cm}^2 \lesssim \sigma_{\chi e} \lesssim 3\times10^{-28}~\mathrm{cm}^2\)] is excluded at 2\(\sigma\) for $m_\chi\lesssim 10^{-5}$ GeV from XENONnT (PandaX) [LZ].
Other constraints are also shown in Fig.~\ref{fig:mchi_sigmae_pair} for comparison. Note that since the PBHBDM is only a subdominant DM component, these traditional boosted DM constraints may not apply here.

The right panel of Fig.~\ref{fig:mchi_sigmae_pair} shows the corresponding 2\(\sigma\) exclusions in the plane of $\sigma_{\chi n}^{\rm SI}$ vs. $m_\chi$. The lower limit is obtained from the minimum signal rate required for exclusion, while the upper limit is caused by attenuation effect that suppresses the detectable flux. In contrast to the DM--electron scattering, the dependence on $m_\chi$ is weaker here because the nuclear target is much heavier. Thus, the characteristic kinematic features are largely washed out, which makes the exclusion contour nearly flat for $m_\chi<10^{-2}$ GeV. We find the region of \(3.8\times10^{-38}~\mathrm{cm}^2 \lesssim \sigma_{\chi n}^{\rm SI} \lesssim 4.5\times10^{-27}~\mathrm{cm}^2\) (\(2.2\times10^{-37}~\mathrm{cm}^2 \lesssim \sigma_{\chi n}^{\rm SI} \lesssim 1.8\times10^{-27}~\mathrm{cm}^2\)) [\(9.5\times10^{-38}~\mathrm{cm}^2 \lesssim \sigma_{\chi n}^{\rm SI} \lesssim 4.5\times10^{-27}~\mathrm{cm}^2\)] is excluded at 2\(\sigma\) from XENONnT (PandaX) [LZ].

In addition, we find that although the deeper location of PandaX improves the rejection of certain backgrounds, it also increases the attenuation effect for the PBHBDM signal. As a result, the upper limit is slightly weaker. Among the three experiments, XENONnT yields the most stringent exclusions in both cases because its best-fit background results into a better agreement with the measured data with a smaller $\chi^2_0$.

\subsection{Constraints on the fraction of PBHs as DM today}

\begin{figure}[htbp]
\centering
\includegraphics[width=0.5\textwidth]{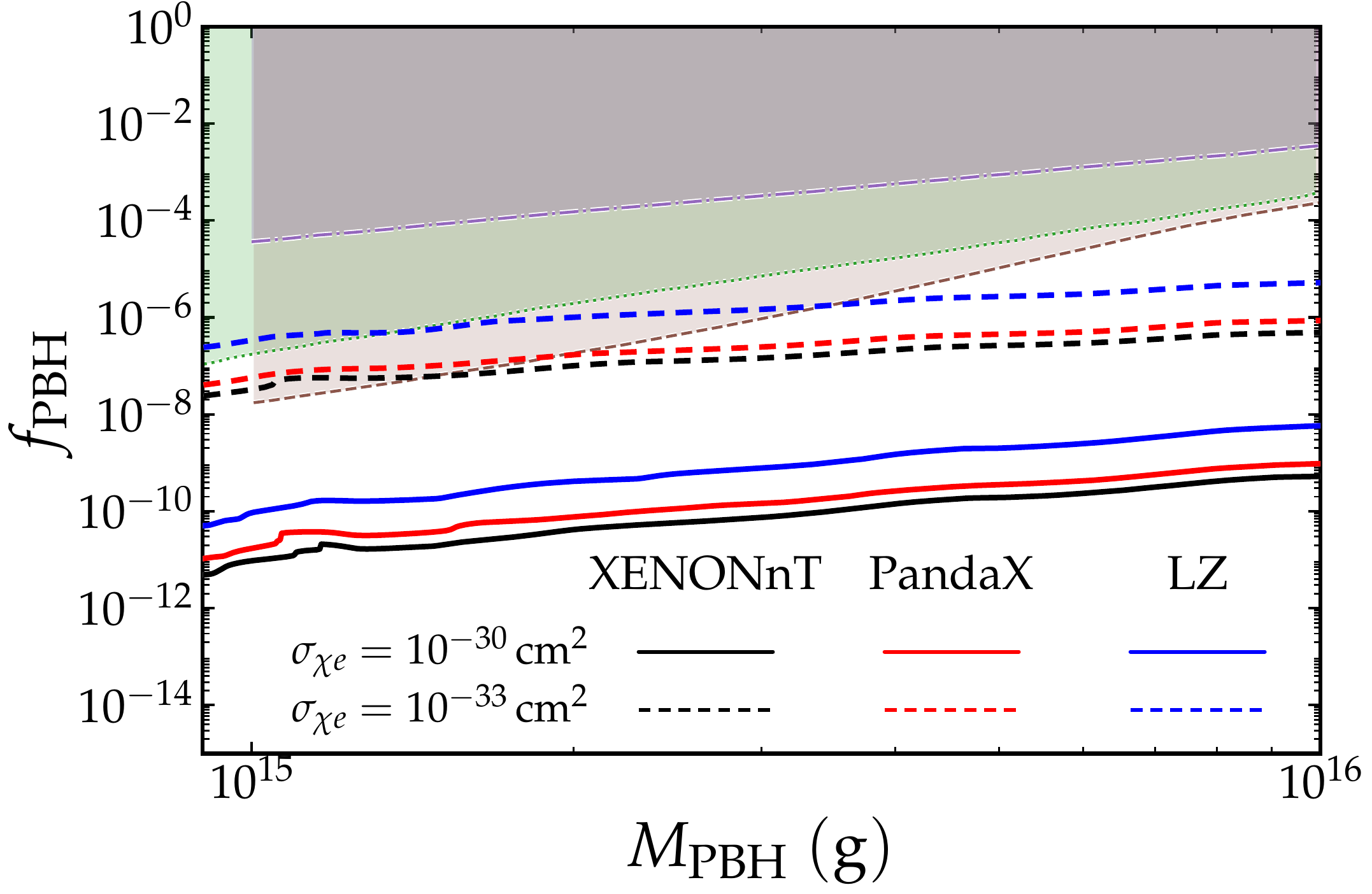}\hfill
\includegraphics[width=0.5\textwidth]{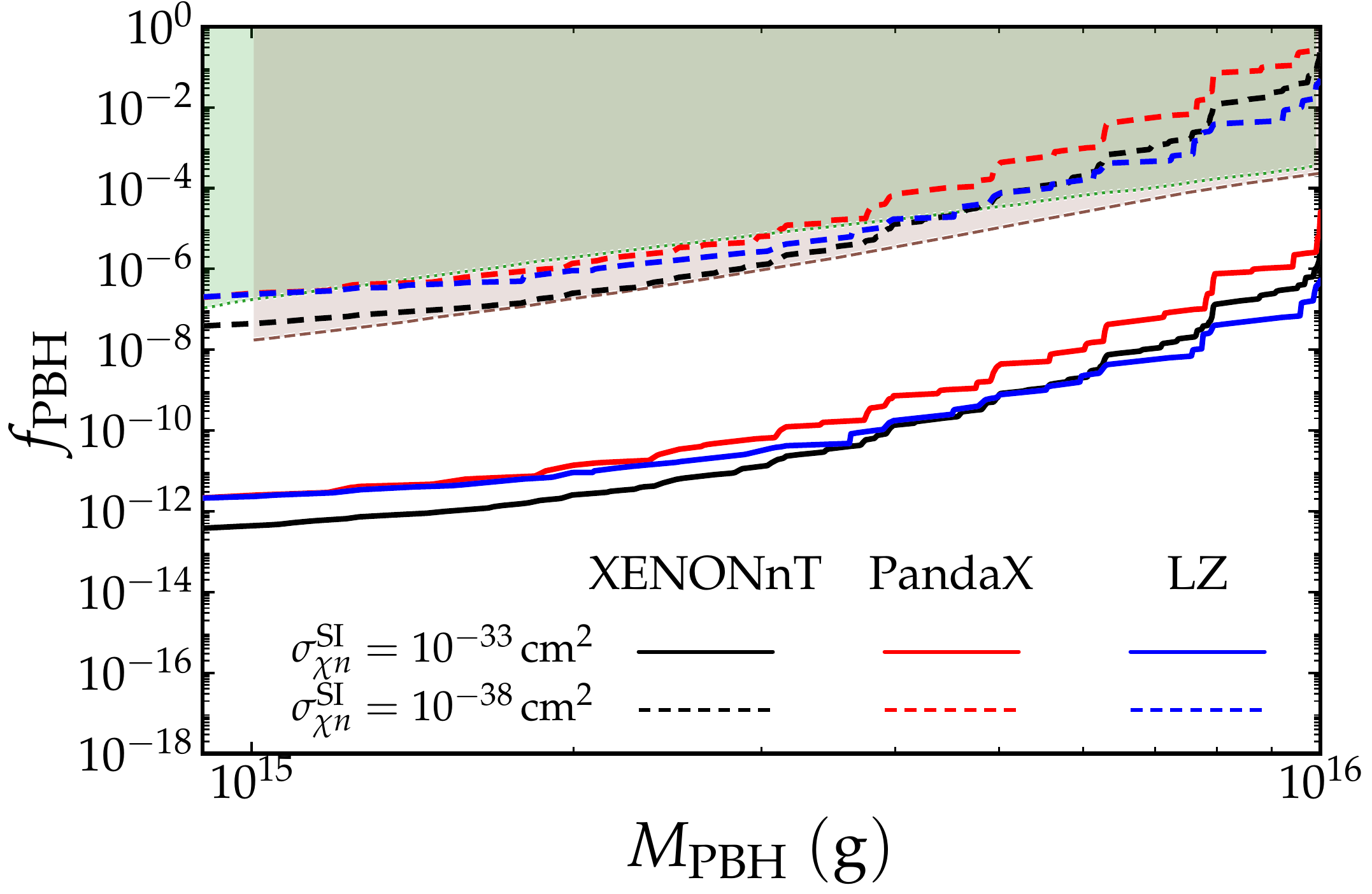}
\caption{Left: \(2\sigma\) bounds on \(f_{\mathrm{PBH}}\) as a function of \(M_{\mathrm{PBH}}\) for
\(\sigma_{\chi e}=10^{-30}\,\mathrm{cm^2}\) (solid lines) and \(\sigma_{\chi e}=10^{-33}\,\mathrm{cm^2}\) (dashed lines) with \(m_{\chi}=1\,\mathrm{MeV}\).
Right: \(2\sigma\) bounds on \(f_{\mathrm{PBH}}\) as a function of \(M_{\mathrm{PBH}}\) for \(\sigma_{\chi n}^{\rm SI}=10^{-33}\,\mathrm{cm^2}\) (solid lines) and \(\sigma_{\chi n}^{\rm SI}=10^{-38}\,\mathrm{cm^2}\) (dashed lines) with Helm form factor and \(m_{\chi}=1\,\mathrm{MeV}\).
The bound are from XENONnT (black), PandaX (red), and LZ (blue). The brown region corresponds to the EDGES 21cm constraint on the PBH abundance from Ref.~\cite{Clark:2018ghm}. The green region is the IGRB constraint without astrophysical component modeling~\cite{Chen:2021ngo}. The purple region in the left panel is the constraint from CDEX-10 experiment with $\sigma_{\chi e}=10^{-29}~{\rm cm}^2$~\cite{CDEX:2024xqm}.
}
\label{fig:fpbh_vs_mpbh_pair}
\end{figure}

Fig.~\ref{fig:fpbh_vs_mpbh_pair} shows the $2\sigma$ upper limits on the fraction of DM composed of PBHs $f_{\rm PBH}$ as a function of the PBH mass $M_{\rm PBH}$ for $m_\chi=1~\mathrm{MeV}$. In the left panel, we fix $\sigma_{\chi e}=10^{-30}~\mathrm{cm}^2$ (solid lines) and $10^{-33}~\mathrm{cm}^2$ (dashed lines) for the DM--electron scattering case, while in the right panel we fix $\sigma_{\chi n}^{\rm SI}=10^{-33}~\mathrm{cm}^2$ (solid lines) and $10^{-38}~\mathrm{cm}^2$ (dashed lines) for the DM--nucleus scattering case. Here, the gray region corresponds to the EDGES 21 cm constraint on $f_{\rm PBH}$~\cite{Clark:2018ghm}.
In both panels, the constraints from DM DD experiments become stronger for a larger cross section and a smaller $M_{\rm PBH}$. The latter comes from the fact that Hawking temperature rises with decreasing $M_{\rm PBH}$, leading to an enhanced production of DM particles. This feature is more apparent in the DM--nucleus scattering case, where the exclusion curves exhibit a steeper change over the displayed range of $M_{\rm PBH}$. Among the three experiments, XENONnT typically yields the most stringent bound. For \(M_{\mathrm{PBH}}=9\times10^{14}\,\mathrm{g}\) and \(\sigma_{\chi e}=10^{-30}\,\mathrm{cm^2}\) ($10^{-33}\,\mathrm{cm^2}$), we find \(f_{\mathrm{PBH}} \lesssim 4.5\times10^{-12}\) ($2.5\times10^{-8}$) from XENONnT constraint. For \(M_{\mathrm{PBH}}=9\times10^{14}\,\mathrm{g}\) and \(\sigma_{\chi n}^{\rm SI}=10^{-33}\,\mathrm{cm^2}\) ($10^{-38}\,\mathrm{cm^2}$), the constraint becomes \(f_{\mathrm{PBH}} \lesssim 3.6\times10^{-13}\) ($3.6\times10^{-8}$).

\section{Mass evolution effects on PBHs}
\label{sec:evolution}

As PBHs evaporate, their masses will decrease, and the spectra of emitted DM particles will be altered, as one can see from Eq.~\eqref{eq:hawking_radiation}. In the previous discussion, we follow the predominant approach in the literature by assuming a monochromatic mass for PBHs and set constraints from the instantaneous emission spectra of DM accordingly. This treatment is reasonable for non-evaporated PBHs with mass $M_{\rm PBH}$ above the evaporation threshold, i.e., $M_{\rm th}\approx 7.5\times10^{14}\,\mathrm{g}$.%
\footnote{Note that for the SM case, we obtain $M_{\rm th}\approx 7.45\times10^{14}\,\mathrm{g}$ by using \textbf{BlackHawk v2.3} with HERWIG hadronization. However, the presence of an additional DM particle will slightly accelerate the PBH evaporation.} The evaporation threshold is defined as the initial mass of PBHs that would have fully evaporated by the present cosmic time. In this work, we adopt an age of the Universe $t_0\approx 4.4\times10^{17}\,\mathrm{s}$, which is consistent with the latest cosmological constraints in Ref.~\cite{Planck:2018vyg}. However, for lighter PBHs that are fully evaporated today, one has to take into account the PBH mass loss and perform an explicit integration over the PBH evolution~\cite{Bernal:2022swt}. In this section, we will apply a method similar to that of Ref.~\cite{Bernal:2022swt} to explore lighter PBH masses for PBHBDM.

For a neutral, non-rotating Schwarzschild black hole, the mass-loss rate is typically given by~\cite{Carr:2020gox}
\begin{equation}
\frac{dM_{10}}{dt} = -5.34 \times 10^{-5} \, \mathcal{F}(M_{\rm PBH}) \, M_{10}^{-2} \, \mathrm{s^{-1}}\,, \quad M_{10} \equiv \frac{M_{\rm PBH}}{10^{10}\,\mathrm{g}}\,,
\label{eq:mass_loss}
\end{equation}
where $\mathcal{F}(M_{\rm PBH})$ is a coefficient quantifying the particle species contributing to the Hawking radiation. It is conventionally normalized to unity for PBHs with masses $M_{\rm PBH} \gg 10^{17}\,\mathrm{g}$, and increases as the mass decreases, reflecting the inclusion of additional particle species as their emission channels open. As a PBH enters its final evaporation phase, the Hawking temperature $T_{\rm PBH}$ increases to a very high temperature such that all SM particles can be kinematically emitted. In this regime, $\mathcal{F}(M_{\rm PBH})$ effectively approaches a constant value, $\mathcal{F}(M_{\rm PBH}) \approx \mathcal{F}_{\rm SM} = 15.35$.
Our setup incorporates an additional DM particle $\chi$ whose emission modifies $\mathcal{F}(M_{\rm PBH})$. The $\chi$ particles provide a constant increment $\Delta \mathcal{F}_\chi$ to the SM emission coefficient $\mathcal{F}_{\rm SM}$. Following the graybody calculations by Page and the standard parameterizations adopted in PBH literature~\cite{Page:1976df,Carr:2020gox}, a neutral spin-${1\over 2}$ Dirac field contributes $\mathcal{F}_{1/2}=0.147$ per internal degree of freedom. Thus, the increment is $\Delta \mathcal{F}_\chi=g_\chi \mathcal{F}_{1/2}=0.588$. These effects have been taken into account by \textbf{BlackHawk}. The emission spectrum is tabulated with each evaporation time, thus the evaporation rate for each $M_{\rm PBH}$ can be readily determined in \textbf{BlackHawk}.

\begin{figure}[htbp]
\centering
\includegraphics[width=0.7\textwidth]{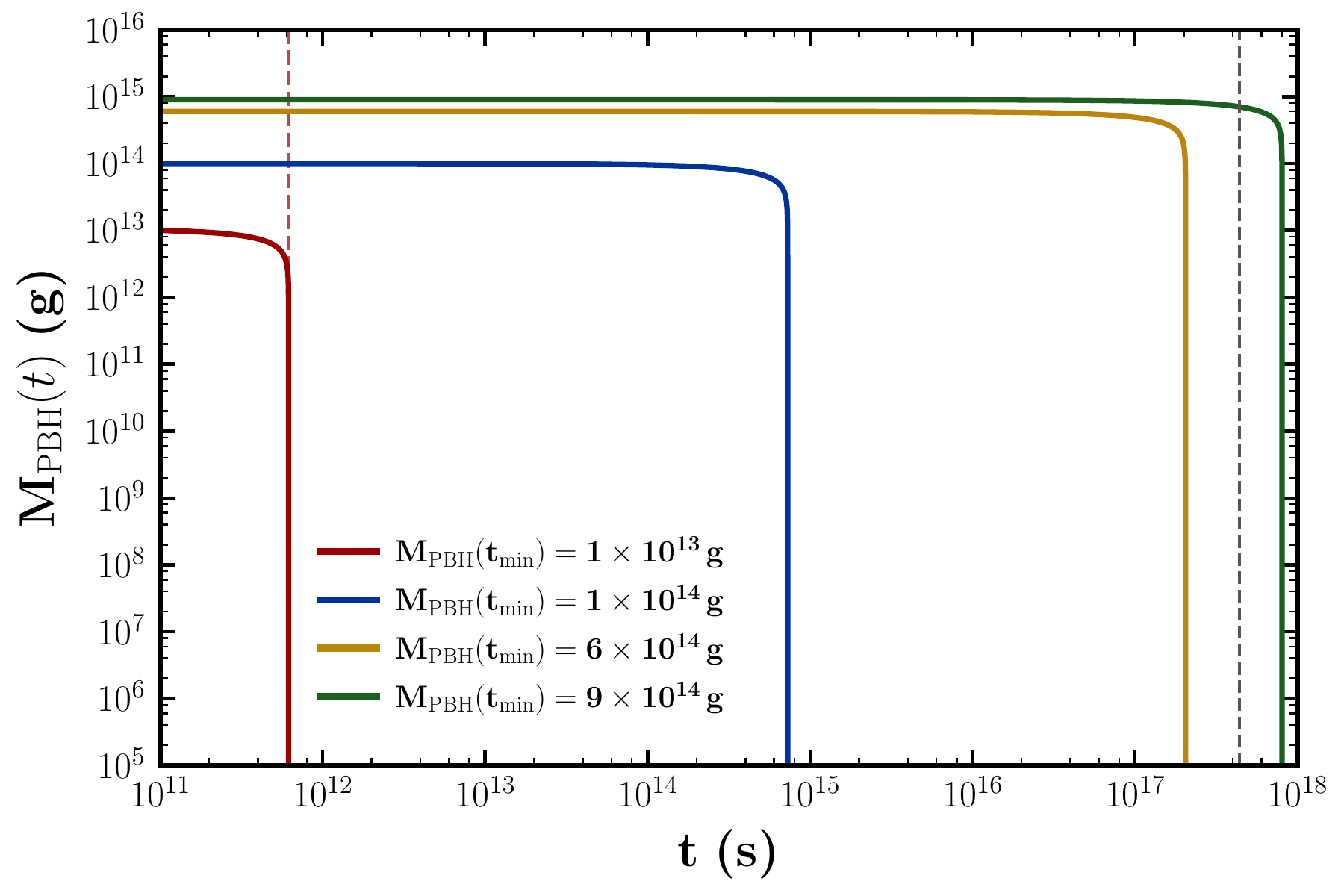}
\caption{
Evolution of the PBH mass $M_{\rm PBH}(t)$ as a function of cosmic time $t$ for four representative initial masses. The solid curves represent $M_{\rm PBH}(t_{\min}) = 1\times10^{13}\,\mathrm{g}$ (red), $1\times10^{14}\,\mathrm{g}$ (blue), $6\times10^{14}\,\mathrm{g}$ (yellow), and $9\times10^{14}\,\mathrm{g}$ (green). The vertical red dashed line mark the epoch where the PBH remaining lifetime is exactly one year ($\Delta t_{\rm obs} = 1\,\mathrm{yr}$) for $M_{\rm PBH}(t_{\min}) = 1\times10^{13}\,\mathrm{g}$, indicating the onset of the final explosive phase. The vertical gray dashed line denotes the current age of the Universe.
}
\label{fig:PBHmass_time}
\end{figure}

Here we define instantaneous mass $M_{\rm PBH}(t)$ to account for the evaporation process, with $M_{\rm PBH}(t_{\min})$ as the initial mass.
The time evolution for $M_{\rm PBH}(t)$ as a function of $t$ is shown in Fig.~\ref{fig:PBHmass_time}. One can see that $M_{\rm PBH}(t)$ remains nearly constant for the majority of its existence, followed by a phase of rapid mass loss as the evaporation enters its terminal stage. 
Thus, when $M_{\rm PBH}(t_{\min}) \gtrsim 9\times10^{14}\,\mathrm{g}$, we can treat the emitted spectra as a constant.
For PBHs with initial masses $M_{\rm PBH}(t_{\min}) < M_{\rm th}$, they completely evaporate at the time $t_{\rm evap}$, which is smaller than $t_0$. We also define a typical exposure time of current DD experiments as $\Delta t_{\rm obs}=1\,\mathrm{yr}\approx 3.16 \times 10^7~\mathrm{s}$.

Based on their mass loss on the cosmic scale, and the exposure time of DM DD experiments, we can divide the PBH mass ranges into three parts:

\begin{enumerate}
\item $M_{\mathrm{PBH}} \gtrsim 9\times10^{14}\,\mathrm{g}$. The total lifetimes of relatively massive PBHs are long enough and their emission spectra remain nearly unchanged throughout their existence. In such case, the PBHs emit Hawking radiation throughout their lifetimes and the nature of radiation is primarily determined by their initial masses. The spectrum at any chosen time closely agrees with the initial emission spectrum at formation. As indicated by the solid green curve in Fig.~\ref{fig:PBHmass_time}, the PBHs with initial mass $M_{\rm PBH}(t_{\min})=9\times10^{14}\,\mathrm{g}$ lose only a negligible amount of mass before $t_0$.
\item $6\times10^{14}\,\mathrm{g} \lesssim M_{\mathrm{PBH}} \lesssim 9\times10^{14}\,\mathrm{g}$. For PBHs with masses close to the threshold $M_{\rm th}$, the PBHs exhibit rapid mass loss in the final stage of evaporation. This leads to significant change in the emitted spectrum with an intense burst of high energy particles~\cite{Klipfel:2025jql}. PBHs in their terminal phase may be situated in close proximity to the observer, potentially exerting a significant impact on the galactic component of the total flux. A more careful treatment of the observation time is required but is beyond the scope of this work. Therefore, we restrict our analysis to a conservative scenario, and do not account for this specific mass interval. As an illustration, we consider a PBH with a distance $r_h$ from the Earth that fully evaporates at cosmic time $t_{\rm evap}^\star$. Its final relativistic burst is observed at the present time $t_0$. The corresponding arrival time relation is $t_0=t_{\rm evap}^\star + r_h/c$, which implies $t_{\rm evap}^\star = t_0 - r_h/c$. By taking $r_h=200~\mathrm{kpc}$ as a representative halo radius, we find $t_{\rm evap}^\star\simeq 4.3998\times 10^{17}~\mathrm{s}$. For comparison, the PBHs with initial mass $M_{\rm PBH}=6\times10^{14}\,\mathrm{g}$ fully evaporate by $t_{\rm evap}\simeq 2.0495\times 10^{17}~\mathrm{s}<t_{\rm evap}^\ast$. They would have completed evaporation well before the epoch required for their final burst to arrive today from the halo edge. Consequently, the PBHs with $M_{\rm PBH}\lesssim 6\times10^{14}\,\mathrm{g}$ evaporate too early to contribute to a present-day MW component.
\item $10^{13}\,\mathrm{g} \lesssim M_{\mathrm{PBH}} \lesssim 6\times10^{14}\,\mathrm{g}$. For PBHs with masses below the threshold $M_{\rm th}$, the previous simplified treatment fails to describe the change of their spectrum and possibly complete evaporation. We therefore deal with the galactic contribution by consistently associating the spectrum with the appropriate time , and perform a full time integral by changing $t_0$ in Eq.~(\ref{eq:eg_flux}) to $M_{\mathrm{PBH}}$ dependent $t_{\rm evap}$ in Eq.~(\ref{eq:eg_flux_evolution}). In this regime, as the evaporation enters its final stage, the PBH mass may change significantly even for observers on Earth. Similar to Sec.~\ref{sec:PBH}, we calculate the diffuse flux reaching Earth by accounting for the mass evolution over cosmic time. However, the observation time is integrated in Sec.~\ref{sec:scattering} for each experiment as exposure and thus the spectrum change of certain short final stage is neglected. This is reasonable as shown by the vertical red dashed line in Fig.~\ref{fig:PBHmass_time} for the PBHs with initial mass $M_{\rm PBH}(t_{\min})=1\times10^{13}\,\mathrm{g}$. Such PBHs fully evaporate by $t_{\rm evap}\simeq 5.17\times 10^{11}~\mathrm{s}\gg \Delta t_{\rm obs}$ and the interval $\Delta t_{\rm obs}$ covers only a negligible fraction of the total evaporation history.  Consequently, for PBHs with $M_{\rm PBH}>1\times10^{13}\,\mathrm{g}$, the impact of this approximation is expected to be even smaller. Furthermore, due to the time dilation effect, the spectral evolution of a distant PBH appears ``slowed down'' to an observer on Earth, this provides additional support to the validity of this approximation. However, for PBHs with significantly lower masses, the runaway mass loss is so rapid that the spectral change during the exposure time would be non-negligible even with the redshift buffer, requiring a more complex treatment.
\end{enumerate}

In this section, we mainly focus on the PBHs with mass in the ranges $10^{13}\,\mathrm{g} \lesssim M_{\mathrm{PBH}} \lesssim 6\times10^{14}\,\mathrm{g}$.
For PBHs with evolving masses, the PBHBDM flux should be calculated in a different way from that in Sec.~\ref{sec:PBH}. We consider a uniform, isotropic astronomical diffuse model that consists of each layer of PBHs associated with the corresponding distance between the emitter and the Earth.
The DM fluxes from outer layers will take a longer time to reach the Earth, and the influence of their mass changes will manifest layer by layer. For fully evaporated PBHs with $M_{\mathrm{PBH}} \lesssim 6\times10^{14}\,\mathrm{g}$, the flux from MW in Eq.~\eqref{eq:mw_flux} will no longer have contribution to the total flux. Thus, the DM flux will be only given by the extragalactic sources~\cite{Bernal:2022swt}
\begin{equation}
\frac{d^2\phi_\chi^{\mathrm{EG}}}{dT_\chi\,d\Omega}
= \frac{n_{\mathrm{PBH}}(t_0)}{4\pi}
\int_{t_{\min}}^{t_{\rm evap}} \!dt\,[1+z(t)]\,
\left.\frac{d^2N_\chi}{dT_\chi\,dt}\right|_{E_\chi^s}\,,
\label{eq:eg_flux_evolution}
\end{equation}
where $t_{\rm evap}$ denotes the time at which a PBH with a chosen mass is fully evaporated. Here, we derive the time $t_{\rm evap}$ by dynamically changing $M_{\mathrm{PBH}}$ throughout the evaporation process in \textbf{BlackHawk}. Since the PBHs that have already evaporated will not constitute the DM's composition today, we cannot use Eq.~\eqref{eq:npbh_comoving_f} to compute the comoving number density of PBHs $n_{\mathrm{PBH}}(t_0)$, which depends on the fraction of DM composed of PBHs today $f_{\rm PBH}$. Thus, for fully evaporated PBHs, we replace $n_{\mathrm{PBH}}(t_0)$ in Eq.~(\ref{eq:eg_flux_evolution}) with the following expression~\cite{Bernal:2022swt}
\begin{align}
n_{\rm PBH}(t_0)
&= \beta'_{\mathrm{PBH}}\;
\frac{(1/{\rm Gpc})^{3}}{7.98\times10^{-29}}\,
\left(\frac{M_\odot}{M_{\rm PBH}}\right)^{3/2} \notag \\
&\simeq
\big(4.27\times10^{-55}\,{\rm cm}^{-3}\big)\;
\beta'_{\rm PBH}\,
\left(\frac{M_\odot}{M_{\rm PBH}}\right)^{3/2}
\,.
\label{eq:npbh_comoving_beta}
\end{align}
Here $M_\odot=1.99\times 10^{33}\ {\rm g}$ is the mass of the Sun, and $\beta'_{\mathrm{PBH}}$ is a rescaled quantity that represents the initial abundance of evaporated PBHs~\cite{Carr:2020gox,Bernal:2022swt}
\begin{equation}
\beta'_{\rm PBH} \equiv \kappa^{1/2}
\left( \frac{g_{\ast i}}{106.75} \right)^{-1/4}
\left( \frac{h}{0.67} \right)^{-2}
\beta(M_{\rm PBH})\,,
\label{eq:beta_prime}
\end{equation}
where $\kappa$ denotes the collapse efficiency representing the fraction of the horizon mass collapses into a PBH, $g_*^i$ is the effective number of relativistic degrees of freedom at formation, $h$ is the Hubble parameter today in unit of 100 km/s/Mpc, and $\beta(M_{\rm PBH})$ is the standard PBH formation fraction with a full expression given in Ref.~\cite{Carr:2020gox}.

\begin{figure}[htbp]
\centering
\includegraphics[width=0.5\textwidth]{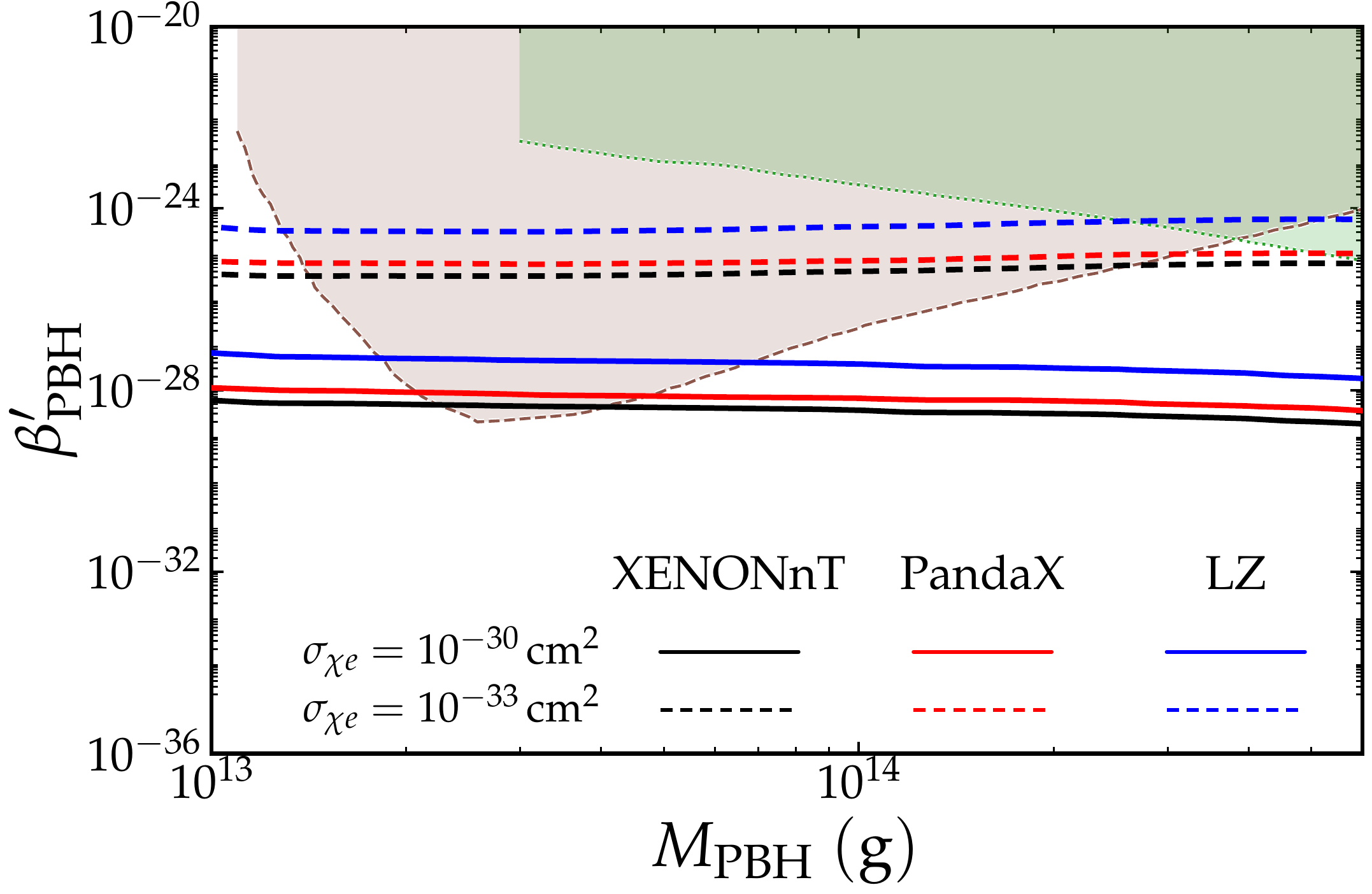}\hfill
\includegraphics[width=0.5\textwidth]{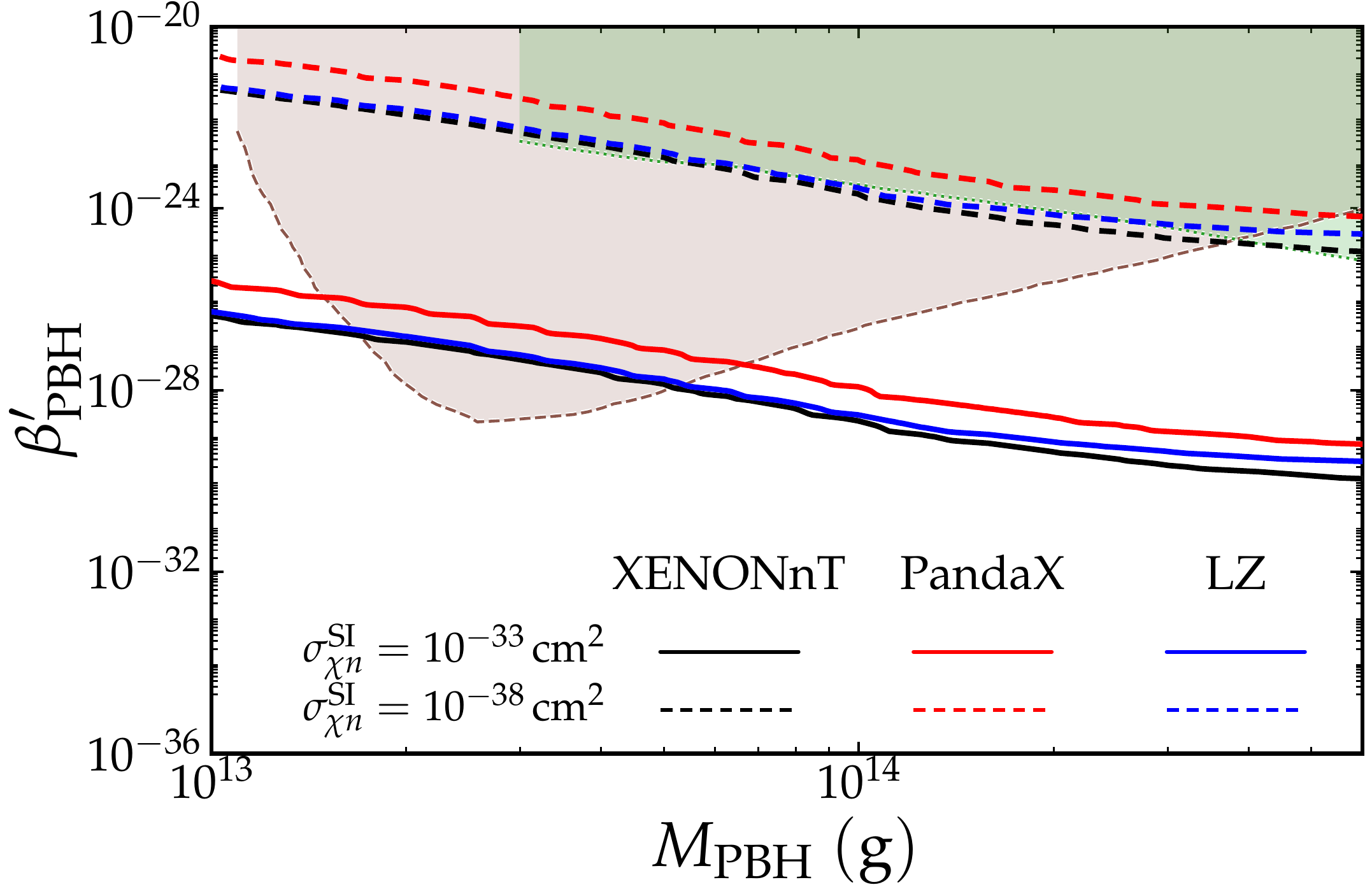}
\caption{Left: \(2\sigma\) bounds on $\beta'_{\rm PBH}$ as a function of \(M_{\mathrm{PBH}}\) for
\(\sigma_{\chi e}=10^{-30}\,\mathrm{cm^2}\) (solid lines) and \(\sigma_{\chi e}=10^{-33}\,\mathrm{cm^2}\) (dashed lines) with \(m_{\chi}=1\,\mathrm{MeV}\).
Right: \(2\sigma\) bounds on $\beta'_{\rm PBH}$ as a function of \(M_{\mathrm{PBH}}\) for \(\sigma_{\chi n}^{\rm SI}=10^{-33}\,\mathrm{cm^2}\) (solid lines) and \(\sigma_{\chi n}^{\rm SI}=10^{-38}\,\mathrm{cm^2}\) (dashed lines) with Helm form factor and \(m_{\chi}=1\,\mathrm{MeV}\).
The bounds are from XENONnT (black), PandaX (red), and LZ (blue). The green region corresponds to the IGRB constraint on the PBH abundance from Ref.~\cite{Chen:2021ngo} and the brown region shows the CMB anisotropy constraint from Ref.~\cite{Acharya:2020jbv}.}
\label{fig:fpbh_vs_mpbh_pair_evolution}
\end{figure}

In Fig.~\ref{fig:fpbh_vs_mpbh_pair_evolution}, after encoding the PBH abundance at formation, we recast the constraints in terms of the parameter \(\beta'_{\mathrm{PBH}}\). As seen from the left panel of Fig.~\ref{fig:fpbh_vs_mpbh_pair_evolution}, the constraints do not weaken as much below the evaporation mass $M_{\rm th} \approx 7.5 \times 10^{14}~\mathrm{g}$. We assume a monochromatic PBH mass function with PBHs formed uniformly at a single cosmic time throughout the Universe. In this setup, when the PBH mass is lower than the evaporation mass, the time integral relevant for the accumulated flux does not extend over the full cosmic history. Consequently, the lighter PBHs within this mass range, e.g., $10^{13}\,\mathrm{g} \le M_{\mathrm{PBH}} \le 6\times10^{14}\,\mathrm{g}$, typically yield a lower observable flux. However, this decreasing effect is cancelled out by the parameter change from $f_{\rm PBH}$ to $\beta'_{\rm PBH}$. Since \(\beta'_{\mathrm{PBH}}\) is correlated with \(M_{\mathrm{PBH}}\), lowering \(M_{\mathrm{PBH}}\) modifies the prefactor of \(n_{\mathrm{PBH}}\), and thus compensates the weakening of the limits as the mass decreases. For \(M_{\mathrm{PBH}}=6\times10^{14}\,\mathrm{g}\) and \(\sigma_{\chi e}=10^{-30}\,\mathrm{cm^2}\) (\(\sigma_{\chi n}^{\rm SI}=10^{-33}\,\mathrm{cm^2}\)), we find \(\beta'_{\mathrm{PBH}} \lesssim 2\times10^{-27}\) (\(\beta'_{\mathrm{PBH}} \lesssim 1\times10^{-30}\)) from the most stringent XENONnT. For \(\sigma_{\chi e}=10^{-33}\,\mathrm{cm^2}\) (\(\sigma_{\chi n}^{\rm SI}=10^{-38}\,\mathrm{cm^2}\)), the corresponding limit relaxes to \(\beta'_{\mathrm{PBH}}\lesssim 8\times10^{-26}\) (\(\beta'_{\mathrm{PBH}}\lesssim 1.2\times10^{-25}\)).

\section{Conclusions}
\label{sec:conclusions}

We have investigated the sensitivities of underground DM DD experiments to light DM produced by Hawking evaporation of PBHs. The attenuation of the DM flux in the Earth is taken into account in details. Using the latest XENONnT, PandaX and LZ data, we derived new constraints on sub-GeV PBHBDM and on the PBH abundance by taking into account both the DM--electron and DM--nucleus scattering. Our main quantitative results are summarized as follows:
\begin{enumerate}
\item For the benchmark PBH parameters \(M_{\mathrm{PBH}} = 10^{15}\,\mathrm{g}\) and \(f_{\mathrm{PBH}} = 1.6\times10^{-8}\), the DM--electron (DM--nucleus) scattering cross section range \(7.8\times10^{-34}~\mathrm{cm}^2 \lesssim \sigma_{\chi e} \lesssim 4.5\times10^{-28}~\mathrm{cm}^2\) (\(3.8\times10^{-38}~\mathrm{cm}^2 \lesssim \sigma_{\chi n}^{\rm SI} \lesssim 4.5\times10^{-27}~\mathrm{cm}^2\)) is excluded at \(2\sigma\). The new data in DM DD experiments greatly improve the sensitivity compared with previous studies.
\item We also impose constraints on the fraction of DM in PBHs \(f_{\mathrm{PBH}}\) for not fully evaporated PBHs in the mass range of \(9\times10^{14}\text{--}1\times10^{16}\,\mathrm{g}\). For a fixed scattering cross section, the bounds on \(f_{\mathrm{PBH}}\) generally become more stringent as \(M_{\mathrm{PBH}}\) decreases. The dependence of limits on $M_{\rm PBH}$ for DM--nucleus scattering is stronger than that for DM--electron scattering.
\item By taking into account the PBH mass evolution, we obtained the constraints on initial PBH abundance parameter \(\beta'_{\mathrm{PBH}}\) for PBHs with masses of \(1\times10^{13}\text{--}6\times10^{14}\,\mathrm{g}\). Over the mass range in which our analysis is applicable, the bounds derived from the DM signal are comparable to, and often more stringent than existing gamma-ray constraints, especially in the DM–electron scattering.
\end{enumerate}

\acknowledgments
T.~L. is supported by the National Natural Science Foundation of China under Grant No.~12375096. J.~L. is supported by the National Natural Science Foundation of China under Grant No.~12275368.

\bibliography{refs}

\end{document}